\documentclass{article}
\usepackage{iclr2027_conference,times}
\usepackage{amsmath,amssymb,amsthm,graphicx,booktabs,array,longtable}
\usepackage{hyperref,url}
\hypersetup{colorlinks=true,linkcolor=blue,citecolor=blue,urlcolor=blue,pdfauthor={Alireza Moghaddasi},pdftitle={Measuring Counterfactual Credit in a Replayable Wildfire-Suppression Benchmark}}
\usepackage{xcolor}
\title{Measuring Counterfactual Credit in a\\Replayable Wildfire-Suppression Benchmark}
\author{Alireza Moghaddasi\\
Department of Civil and Environmental Engineering\\
University of Maryland, College Park, MD, USA\\
{\fontfamily{cmtt}\selectfont moghaddasi.alireza.am@gmail.com}}
\iclrfinalcopy
\begin{document}
\maketitle
\lhead{Preprint}

\begin{abstract}
In cooperative multi-agent reinforcement learning, learned critics are usually evaluated by how well they predict the team's return. Yet predicting  a team's return and assigning credit to individual actions are distinct tasks.  To test credit directly, we introduce FLARE (Fireline Learning Agents in a Replayable Environment), a stochastic wildfire suppression benchmark in which any moment of a fire can be replayed with different actions under the same random events. Replaying one step measures the immediate effect of each agent's suppression, which we call simulator credit and use for training. Replaying to the end of the episode measures the effect of each action, which we use for grading critics. In  a comparison over eight seeds with multi-agent proximal policy optimization (MAPPO), simulator credit reduces the interquartile mean  burn score (each cell's burning time, capped and averaged over the map; lower is better) from $26.6\%$ with shared reward to $20.9\%$.  Learned credit from a centralized COMA-style critic does no better than the shared reward. Four  trained critics predict the team's return well, yet  the credit they assign is 9 to 42 times larger than the credit measured by replay.   The critics also give heterogeneous agents nearly identical action rankings.  These results show that a good fit to the team's return is a weak test of a critic's credit. 
\end{abstract}

\section{Introduction}\label{sec:intro}
In cooperative multi-agent reinforcement learning, a team of agents is often trained with a single reward that every agent shares \citep{papoudakis2021benchmarking,yu2022surprising}. A shared reward tells each agent how the team did, but it does not tell the agent how much its own action contributed, because the same reward follows a useful action, a redundant one and an action with no effect. The credit assignment problem asks how to give each agent feedback that reflects its own contribution, and methods differ in how they estimate it: by value decomposition, which splits the team's value into one part for each agent \citep{sunehag2018vdn,rashid2018qmix}, by Shapley values \citep{wang2020shapley}, or by comparing the outcome with a counterfactual  \citep{foerster2018coma}. A classic counterfactual answer is the difference reward: the team's reward minus the reward the team would have received had the agent not acted \citep{wolpert1999introduction}.

Computing a difference reward requires the team's outcome in a situation that did not happen: the same moment without the agent's action. Tumer and Agogino obtained this counterfactual by running their air traffic simulator once more for each agent \citep{tumer2007distributed}. Such repeated simulation is often unavailable, because it needs a simulator that can be restored to any moment, and it costs an extra simulation for each agent. Deep multi-agent methods therefore learn the counterfactual from experience.

The standard example of this approach is counterfactual multi-agent policy gradients \citep[COMA;][]{foerster2018coma}. It trains a centralized critic that predicts the team's return for each action an agent could take. The agent's credit, which COMA calls its counterfactual advantage, is the predicted return of the action it chose minus the average prediction over all its actions, weighted by how likely the agent is to choose each one. COMA introduced this critic as a cheaper substitute for the extra simulations that difference rewards require. Learned critics are usually judged by how well they predict the team's return or by how well the agents they train perform \citep{castellini2022difference,papoudakis2021benchmarking}. Neither check shows whether the credit given to each of an agent's actions is correct.

In stochastic tasks, random events can move the team's return far more than one agent's choice of action does. A critic can then predict returns well and still give the wrong credit, so the usual checks leave a gap in which large errors in credit can go unnoticed \citep{ilyas2020closer,castellini2022difference}. Checking the credit directly requires the outcome of each alternative action from the same situation, and these outcomes are far more precise when all alternatives share the same random events, a technique known as common random numbers \citep{glasserman1992common}. We therefore ask two questions in a simulator that can replay any situation in this way. Do difference rewards computed by replay help agents learn more than the shared reward or a local reward for suppression next to the fire? Does the credit from a trained COMA-style critic match the credit that replay measures?

FLARE (Fireline Learning Agents in a Replayable Environment) is a wildfire suppression simulator built for these questions. Any moment of a fire can be saved, together with the state of its random number generator, and replayed with different actions under common random numbers, as policy search methods do with resettable simulators \citep{schulman2015trpo}. At every step, eight agents of three types (hand crews, dozers and aircraft) each choose one of 20 actions, a move combined with a suppression choice, and the rules link their actions.

\begin{figure}[t]
\centering
\includegraphics[width=\linewidth]{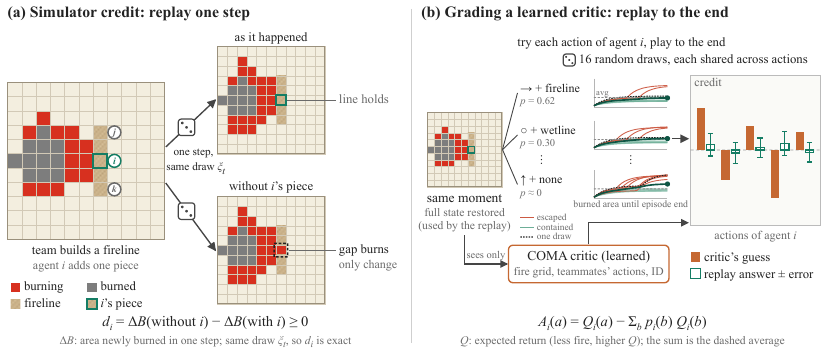}
\caption[Two uses of replay.]{Two uses of replay.  (a) Simulator credit: replay one step with and without one agent's accepted suppression under the same random draw $\xi_t$; the burning that the suppression prevents, normalized by map size, is that agent's credit $d_i$ (Equation~\ref{eq:placement}). (b) Grading a learned critic: replay each of the agent's twenty actions to the end of the episode under sixteen random draws, each shared by all twenty actions, and compare the resulting replay credit (Equation~\ref{eq:main-target}) with the critic's credit. All grids, paths and bars are schematic.}
\label{fig:replay-overview}
\end{figure}

We use replay in two ways (Figure~\ref{fig:replay-overview}). For training, each agent receives a difference reward computed by replaying one step without the suppression that agent applied. This reward credits only the suppression part of an action, so a move by itself earns nothing at that step. The learning algorithm, however, adds the rewards of later steps to each action's return, so a move that brings an agent to the fire still receives credit for the suppression that follows. For evaluation, we take one agent at a saved moment, replay each of its possible actions to the end of the episode many times, and compare the credit that these replays measure with the credit that a trained critic assigns to the same actions.

The training comparison holds the learning algorithm fixed, multi-agent proximal policy optimization \citep[MAPPO;][]{yu2022surprising}, and changes only the reward. With difference rewards, the final burn score (each cell's burning time, capped and averaged over the map; lower is better) falls from 26.6\% under the shared reward to 20.9\%, and difference rewards beat both the shared reward and a local reward for suppression next to the fire in all eight pairs of runs that share a seed.

We also test whether this gain depends on two rules.  In the current rules, which we use by default, ground crews (hand crews and dozers) apply wetline, water on one cell per action, without carrying water, and a fireline, a strip cleared to bare soil, must start from the map edge or from another fireline. The gain remains when ground crews must carry and refill their water and when firelines may also start from burned ground. With both changes, difference rewards give a burn score about ten percentage points lower than the shared reward.

 Replacing MAPPO's usual advantage with the credit of a COMA-style critic does no better than the shared reward. Three attempts to improve it left this result unchanged: removing the clamp that limits the policy's logits to $[-10,10]$, training the critic on replayed outcomes of untaken actions over one step, and, in a small trial, giving each agent type its own output head. Together, these results confirm an earlier finding that exact difference rewards outperform COMA \citep{castellini2022difference}, now with the learning algorithm held fixed. Feedback for each agent is therefore not enough by itself: the COMA-style critic also gives each agent its own credit yet does no better than the shared reward, and the local reward, also given to each agent, loses to difference rewards in all eight seed pairs.

 We then grade the critics from four of these training runs. They pass a standard value check, predicting the team's return with $R^2$ of about 0.9. Yet the credit they assign to individual actions is 9 to 42 times larger than the credit measured by replay. The ratio grows when each action is replayed more often, because noise in the replay can only make the replay credit look larger. These ratios therefore understate the critics' overestimate.

Each critic receives the agent's identity and type, yet it ranks the 20 possible actions almost identically for hand crews, dozers and aircraft, as it already did before training. A short proposition in Section~\ref{sec:theory-summary} formalizes a known argument about extrapolation error \citep{fujimoto2019bcq}: fitting the returns of the actions taken leaves the values of untaken actions, and therefore their credit, undetermined. Replay has a limit of its own. Even 256 replays per action, which we ran for one critic, leave the estimate too noisy to grade the order of an agent's actions, so we grade the size of the credit and report the order as unresolved.

Because difference rewards and common random numbers are established tools, our contributions are a benchmark and two measurements:
\begin{itemize}
\item FLARE, a replayable wildfire suppression benchmark that provides exact difference rewards over one step for training and replay estimates to the end of the episode for evaluation.
\item A comparison of rewards with the learner held fixed, under the current rules and under the two rule changes, alone and together.
\item An evaluation of trained COMA-style critics against replay to the end of the episode. Earlier work graded the accuracy of COMA's value predictions \citep{castellini2022difference}; we grade the size of the credit that each action receives.
\end{itemize}

\section{Related work}
\label{sec:related}
Besides the repeated simulation described in the introduction, difference rewards \citep{wolpert1999introduction} have been computed from a known formula for the team's score and from learned estimates of that score. Coordinated Learning without Exploratory Action Noise (CLEAN) removes the noise that exploring teammates add to each other's rewards \citep{holmesparker2014clean}. Each agent always takes its best current action, so its teammates see no exploration noise.  It explores by calculation instead, using the formula to compute the team's score had it acted differently while its teammates kept their actions. Because such a formula is often unavailable, Colby et al. replaced it with each agent's learned estimate of the team's score and judged the estimate by the team's final performance \citep{colby2015counterfactual}. In tightly coupled tasks, which pay off only when several agents act together, removing one agent often changes nothing, so D++ adds imagined copies of the agent and measures how the team's score changes \citep{rahmattalabi2016dpp}.  FLARE computes the same difference reward by repeated simulation, as Tumer and Agogino did. Because its fire spreads randomly, it replays one step with and without the agent's suppression under the same random draw, so the difference comes from the agent's action alone. 

Many deep methods learn credit from experience, by value decomposition or by training a critic \citep{wang2021qplex,wang2021dop,zhao2025maca}. Studies of learned critics mostly examine their value predictions or the variance and bias of the learning signals they produce. A value network can fit its training targets and still predict the true return poorly \citep{ilyas2020closer}, and centralized critics in multi-agent learning add their own sources of variance and bias \citep{kuba2021variance,lyu2023centralized}. Castellini et al. went further and compared a trained COMA critic's action values with averages over 100 rollouts that follow the policy \citep{castellini2022difference}. Credit, however, is a difference between values, and such differences can be much smaller than the errors in the values, so good value predictions do not guarantee correct credit. Because credit averages over all of an agent's actions, it also depends on actions the agent rarely takes, and the critic sees little data on them \citep{fujimoto2019bcq}. We grade this credit directly, action by action, against replay to the end of the episode, in which all actions of one agent share the same random draws.

Teams of language model agents have recently used exact counterfactuals from restored states, because the text written so far determines the state. Chen et al. restore these text histories to measure each agent's counterfactual credit under a frozen policy, and they report gains over baselines with learned critics \citep{chen2026exact}. Related methods study agents that act in turn \citep{deshmukh2026cosac} or compute an agent's reward by removing it from the team \citep{li2026ccpo}. When the simulator cannot be restored, counterfactual outcomes must be inferred instead. Hindsight information can separate an action's effect from luck \citep{mesnard2021counterfactual}, and structural causal models can infer counterfactual outcomes from logged experience \citep{buesing2019woulda,oberst2019counterfactual}. FLARE instead restores its simulator, random number generator included, to the saved moment and obtains the counterfactual outcomes directly. 

Reinforcement learning for wildfire suppression already includes decentralized control of firefighting aerial robots \citep{haksar2018wildfire} and fast stochastic simulation on graphics processors \citep{cakir2025jaxwildfire}. The free-range-zoo suite includes stochastic multi-agent wildfire suppression \citep{patino2025moasei}, and CREW-Wildfire evaluates teams of language model agents in wildfire response \citep{hyun2025crew}. FLARE is a benchmark built for measuring credit, in which the same replay supplies the difference rewards used in training and the reference used to grade trained critics.

\section{FLARE: Fireline Learning Agents in a Replayable Environment}
\label{sec:benchmark}
FLARE is a cooperative wildfire suppression task, formally a decentralized partially observable Markov decision process (Dec-POMDP), in which fire spreads across a grid according to a stochastic cellular automaton. At every step, eight agents act at the same time: each chooses a move and a suppression action from its own partial view, and the rules decide which suppressions take effect. All agents share one reward, the negative of the burn added in that step.

\paragraph{Task and state.}
The main task uses a $64\times64$ grid, four hand crews, two dozers and two aircraft. A randomly located $5\times5$ ignition starts each episode, which lasts at most 100 steps and ends earlier if no active fire or usable fuel remains. The automaton tracks burning, cumulative burning, fuel and mitigation. 

For $N$ agents, the full state $s_t=(x_t,\tau_t)$ holds the fire grid $x_t$ and tactical state $\tau_t$: positions, types, water, refill and no-fly timers, previous moves, activity and time. Agent $i$ sees only part of this state, $o_{it}=O_i(s_t)$, and acts according to
\begin{equation}
 h_{it}=(o_{i0},a_{i0},\ldots,a_{i,t-1},o_{it}),\qquad
 \pi(\mathbf a_t\mid\mathbf h_t)={\textstyle\prod_{i=1}^N}\pi_i(a_{it}\mid h_{it}).
 \label{eq:task-policy}
\end{equation}
Here $h_{it}$ is its local history and bold symbols collect agents; the product means that each agent decides on its own. In our implementation, agents share one network and see only the current observation: a $21\times21$ window of the map centered on the agent, a coarse summary of the whole fire, its type, and aircraft water and timers, but no teammate positions or resources.

\paragraph{Actions and rules.}
Each action $a_{it}=(m_{it},u_{it})\in\mathcal A$ combines five movements (stay, up, down, left, right) and four suppression choices: none, the type's main suppression action, one-cell wetline or one-cell scratchline, giving $|\mathcal A|=20$. In each step, agents first move, the rules then accept or reject each requested suppression, the accepted ones take effect together, and finally the fire spreads. A hand crew's fireline removes the fuel from one cell and a dozer's from up to three cells, and a fireline must start from the map edge or from another line. Aircraft drops wet up to $3\times3$ cells for one step and require water, inactive refill/no-fly timers, and no ground unit within one cell, diagonals included. No-fly periods are random intervals when aerial suppression is blocked.

 Every aerial suppression action must pass these four checks and uses one unit of water when accepted. Ground one-cell wetline and scratchline remain unanchored direct attack, which need not start from the map edge or a fireline: wetline wets one cell for one step, and scratchline halves its fuel. Ground wetline has no modeled water cost and extinguishes burning at its cell. Several earlier comparisons and controls use earlier rules in which aerial one-cell wetline and scratchline bypassed these checks and used no water (Appendix~\ref{app:experiment-inventory}). Detailed timing, footprints and parameters appear in Appendix~\ref{app:config}.

\paragraph{Objective and replay.}
Let $b_t(c)$ be the burning time accumulated by cell $c$ up to step $t$, capped, and let $B_t=\sum_c b_t(c)$ sum it over the $C=64^2$ cells. All agents receive the same environment reward, and every method is evaluated by
\begin{equation}
 r_t=-\frac{B_{t+1}-B_t}{C},\qquad \text{}\operatorname{BurnScore}=\frac{B_T}{C}.
 \label{eq:benchmark-reward}
\end{equation}
 Because the rules link the agents' actions, for example a crew's move can block an aircraft's drop, a replay must restore the agents' positions, water and timers as well as the fire. To replay a moment, we save $z_t=(s_t,\mathbf h_t)$ together with the code's internal records and the state of the random number generator. If all random draws $\omega$ are fixed, for fire spread, no-fly periods and the agents' sampled actions, the rest of the episode is fully determined. Replays of the same moment with different actions therefore share the same random draws (\emph{common random numbers}), while each agent still reacts to what it observes in its own replay. Sections~\ref{sec:method}--\ref{sec:theory-summary} define these interventions precisely.

\section{Credit signals and policy learning}
\label{sec:method}

We train MAPPO with four feedback signals: the shared reward, simulator credit (the difference reward computed by replay), a local reward, and learned credit (the credit of a COMA-style critic). All four training arms share the actor architecture, Adam actor-optimizer settings, budget and logit clamp. Comparing the shared reward with simulator credit therefore changes only the reward, while learned credit also changes the critic and how the advantage is computed (Table~\ref{tab:credit-signals}).

\paragraph{Shared reward.}
Every agent receives $r^{\rm sh}_{it}=r_t$, the same value estimate $V(x_t)$ and end-of-episode flag. Thus pre-normalization generalized advantage estimation (GAE) gives $\widehat A^{\rm sh}_{it}=\widehat A^{\rm sh}_{jt}$ \citep{schulman2016gae}.

\begin{table}[!ht]
\centering
\small
\setlength{\tabcolsep}{4pt}
\renewcommand{\arraystretch}{1.08}
\begin{tabular}{@{}>{\raggedright\arraybackslash}p{.14\linewidth}>{\raggedright\arraybackslash}p{.27\linewidth}>{\raggedright\arraybackslash}p{.24\linewidth}>{\raggedright\arraybackslash}p{.28\linewidth}@{}}
\toprule
Signal & What each agent receives & Exact or learned? & What the critic sees \\
\midrule
Shared reward & The team reward, the same for every agent & Exact, no replay & Fire grid \\
Simulator credit (difference reward) & Team reward with its placement minus team reward without it (Equation~\ref{eq:placement}) & Exact, one replayed step & Fire grid \\
Local reward & Its accepted cells on or next to burning cells, counted & Exact, no replay & Fire grid \\
Learned credit (COMA-style critic) & Critic's value of its action minus the average over its actions, weighted by its policy (Equation~\ref{eq:learned-credit}) & Learned, no replay & Fire grid, teammates' actions and identity; informed critics also see positions, water and timers \\
\bottomrule
\end{tabular}
\caption[The four feedback signals: what each agent receives, whether the signal is exact or learned, and what the critic sees.]{The four feedback signals. The first three enter MAPPO's returns and generalized advantage estimation with a learned value baseline; learned credit replaces the advantage.}
\label{tab:credit-signals}
\end{table}

\paragraph{Simulator credit.}
Simulator credit is the difference reward for one step \citep{wolpert1999introduction,tumer2007distributed}, computed by replay. Let $P_t=(\ell_{1t},\ldots,\ell_{Nt})$ collect accepted placements, the cells each agent treats in this step, with empty entries for rejected requests, and let $P_t^{-i}$ delete agent $i$'s placement. Let $F$ apply these treatments and advance the fire one step using $\xi_t$, the random draw for fire spread at step $t$ (part of $\omega$). With Section~\ref{sec:benchmark}'s grid $x_t$, burned total $B$ and cell count $C$, define
\begin{align}
 r(x,P,\xi)&=-\{B(F(x,P,\xi))-B(x)\}/C,\nonumber\\
 d_{it}^{\rm place}&=r(x_t,P_t,\xi_t)-r(x_t,P_t^{-i},\xi_t).
 \label{eq:placement}
\end{align}
Thus $d_{it}^{\rm place}$ is the burn that agent $i$'s placement prevents in this step under the same random draw, and it is never negative, up to roundoff, because treatments cannot increase burning. Movement, acceptance decisions, other agents' placements and the fire draw stay fixed. An agent earns zero when it treats no cell, or treats the same cells in the same way as a teammate, because repeating a treatment adds nothing; a partial overlap can still earn credit.  

\paragraph{Local reward.}
The local reward counts the distinct cells an agent treats that are on or next to a cell burning before the step, diagonals included, divided by $C=64^2$; it enters MAPPO's returns and GAE without any replay. Overlapping agents each receive their own count.

\paragraph{Learned credit.}
The base critic receives $\chi_{it}=(x_t,\mathbf a_{-i,t},i)$: fire grid, teammate action codes and the agent's identity. Let $q_\phi(\chi_{it},b)$ be its value prediction after undoing target normalization, in reward-standardized training units. Its COMA-style credit, its counterfactual advantage, is
\begin{equation}
 A^c_{it}=q_\phi(\chi_{it},a_{it})
 -\sum_{b'\in\mathcal A}\pi_i(b'\mid h_{it})q_\phi(\chi_{it},b').
 \label{eq:learned-credit}
\end{equation}
The critic is trained only on an estimate of the team's return after the action actually taken. At that moment it receives no outcome for the other 19 actions, which it can learn about only from other moments in which they were chosen and through shared network weights. Informed critics add positions, resources, timers and other tactical features.

\paragraph{Training settings.}
The three arms that change only the reward use GAE and a fire-grid value critic. Learned credit supplies $A^c$ to PPO \citep{foerster2018coma,schulman2017proximal}. Training uses reward standardization and PopArt-style target normalization \citep{vanhasselt2016popart}. Actors clamp raw action scores (logits) to $[-10,10]$ before forming probabilities. We call a decision \emph{fully saturated} when every raw logit lies strictly outside these bounds: its local policy and entropy gradients vanish. An actor is \emph{saturated} when every decision on its checked final-policy trajectories is fully saturated. 

\section{Grading learned credit with a replay reference}
\label{sec:theory-summary}
 This section defines how we grade learned credit, the second use of replay in Figure~\ref{fig:replay-overview}b. For one agent at a saved moment, we compare the credit that a trained COMA-style critic gives each of its 20 actions (Equation~\ref{eq:learned-credit}) with the credit measured by replaying each action to the end of the episode. We call this comparison the audit. It asks whether a critic that predicts team returns well also gives credit of the right size (calibrated) that ranks each agent's actions in the right order (selective).

\paragraph{Target and reference.}
 Fix a saved moment $z_t$ from Section~\ref{sec:benchmark}, together with the joint action $\mathbf a_t$, the frozen policy $\pi$ and the remaining horizon $H$, which runs to the end of the episode. For training discount $\gamma$, write $\mathbf a_t^{i\leftarrow b}=(\mathbf a_{-i,t},b)$, $p_i(b)=\pi_i(b\mid h_{it})$ and $G_H^\pi(z,\mathbf a;\omega)=\sum_{\ell=0}^{H-1}\gamma^\ell r_{t+\ell}$, with zero rewards after termination. Then
\begin{equation}
 Q_i^\pi(b)=\mathbb E_\omega G_H^\pi(z_t,\mathbf a_t^{i\leftarrow b};\omega),\qquad
 A_i^\pi(b)=Q_i^\pi(b)-\sum_{b'\in\mathcal A} p_i(b')Q_i^\pi(b').
 \label{eq:main-target}
\end{equation}
 In words, the target credit of action $b$ is how much higher the team's expected return is when agent $i$ takes $b$ than when it draws its action from its own policy. It has the form of Equation~\eqref{eq:learned-credit}, with the critic's predictions replaced by expected returns given the complete saved moment. We replay all twenty alternatives with teammates' current actions fixed, re-apply the rules, and follow each branch's observations. Teammates' actions are fixed only at step $t$; after that, every agent follows the frozen policy, using what it observes in its own branch. For $M=16$ independent replays per action, each sharing environmental and policy draws across alternatives, define
\begin{equation}
 Z_{i,m}(b)=G_H^\pi(z_t,\mathbf a_t^{i\leftarrow b};\omega_m)
 -\sum_{b'\in\mathcal A} p_i(b')G_H^\pi(z_t,\mathbf a_t^{i\leftarrow b'};\omega_m),\qquad
 \widehat A_i(b)=\frac1M\sum_m Z_{i,m}(b).
 \label{eq:audit-replay-reference}
\end{equation}
Centering each replay cancels return variation common to all actions. The standard Monte Carlo identities give unbiased estimates with variance proportional to $1/M$ under the specified replay distributions (Appendix~\ref{sec:matched-credit-theory}); their Monte Carlo standard errors use the sample variance of $Z_{i,m}$.

\paragraph{Why factual fit is insufficient.}
 The short proposition promised in Section~\ref{sec:intro} (proof in Appendix~\ref{sec:matched-credit-theory}) states that, for fixed regression targets and an unrestricted function class, fitting the returns of the actions taken leaves the values of untaken actions, and therefore their credit, undetermined. The audit measures the credit that parameter-sharing, bootstrapped neural critics learn in practice.

\paragraph{Grading.}
The saved normalization statistics (PopArt and reward standardization) convert critic predictions to raw reward units. Here $A_i^c(b)$ denotes the resulting raw-scale credit for alternative $b$. \emph{Factual} $R^2$ and root mean squared error (RMSE) compare predictions for actions actually taken with their mean replayed returns. Credit error and its post hoc difference from a zero-credit reference are
\begin{align}
 \mathrm{MSE}_{\mathrm{credit},i}&=\sum_b p_i(b)[A_i^c(b)-\widehat A_i(b)]^2,
 \label{eq:audit-credit-mse}\\
 \Delta_i&=\sum_b p_i(b)\{[A_i^c(b)-\widehat A_i(b)]^2-\widehat A_i(b)^2\}.
 \label{eq:audit-zero-reference}
\end{align}
 Positive $\Delta_i$ means that the critic's raw credit error is larger than the error of giving every action zero credit. We average agents within each moment, then moments, separately by checkpoint, and repeat with uniform action weights. Appendix~\ref{app:audit} reports whole-episode bootstrap intervals and separate Monte Carlo errors.

\paragraph{Calibration and direction.}
 Because PPO normalizes advantages within each batch, credit equal to the replay credit times one large constant would leave PPO's signal unchanged but give a large error in Equation~\eqref{eq:audit-credit-mse}, so we also grade credit with its overall scale removed.  Let $W$ average moments and agents with saved-policy action weights, $R_c^2=W[(A^c)^2]$, $R_A^2=W[\widehat A^2]$ and $\rho_{c,A}=W[A^c\widehat A]/(R_cR_A)$. Here $R_c$ and $R_A$ are the root mean squares of the critic credit and the replay credit, unrelated to the factual $R^2$, and $\rho_{c,A}$ is their cosine similarity. Rescaling the critic's credit to the replay estimate's root mean square, $\widetilde A^c=(R_A/R_c)A^c$, gives
\begin{equation}
 \frac{W[(\widetilde A^c-\widehat A)^2]}{W[\widehat A^2]}=2(1-\rho_{c,A}).
 \label{eq:rms-credit-error}
\end{equation}
  Ranks and split-half reliability checks describe action structure. The check of gradient direction asks whether the critic credit and the replay credit would push the frozen policy in the same direction (Appendix~\ref{app:scale-sensitivity}).  Because a reference built from sixteen replays has a signal-to-noise ratio near one, meaning that its root mean square is about equal to its Monte Carlo error (Section~\ref{sec:audit-results}), we use it to grade the size of the credit and report the order of actions as unresolved. Checks computed from the critic alone, such as comparing its rankings across agents, need no reference.

\section{Experiments and results}
\label{sec:results}
\subsection{Setup}
\label{sec:protocol}
 We train MAPPO with each of the four feedback signals, using informed critics for learned credit, with eight seeds per signal (1 to 8) and 732 updates per run (2,998,272 agent steps). After training, each frozen policy takes its most likely action on the same thirty new fires, none of which was used in training. We average each policy's fire outcomes and report the interquartile mean (IQM) of the eight policy scores \citep{agarwal2021precipice}. The $95\%$ intervals  come from 10,000 bootstrap resamples of training seeds, paired for comparisons. Appendix~\ref{app:config} gives training settings and Appendix~\ref{app:performance} gives evaluation and sensitivity checks.

\subsection{Does simulator credit improve learning?}
\label{sec:learning-results}
\label{sec:repair}
\begin{figure}[htbp]
\centering
\includegraphics[width=\linewidth]{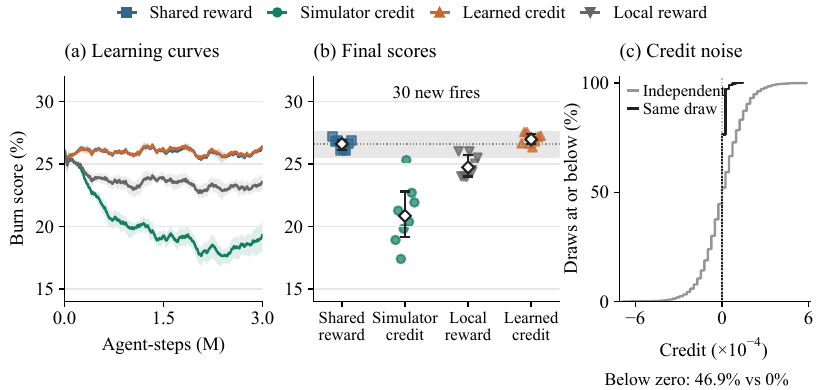}
\caption[Simulator credit improves suppression.]{Simulator credit improves suppression.  (a) Training curves: mean over eight seeds with $95\%$ intervals. (b) Final burn score on thirty new fires: one dot per seed, the interquartile mean with its $95\%$ interval (diamonds), and a gray band of $\pm1$ point around the shared reward. (c) Distribution of simulator credit with independent and with shared random draws, under earlier rules (Section~\ref{sec:ablations}).}
\label{fig:repaired-learning}
\end{figure}

\textbf{Yes, in matched MAPPO training.}  Simulator credit reduces the IQM burn score from $26.6\%$ to $20.9\%$, winning all eight seed pairs (Figure~\ref{fig:repaired-learning}). The difference is $5.76$ percentage points ($95\%$ interval $[3.65,7.73]$), a $21.7\%$ relative improvement.

\textbf{A simple local reward helps, but simulator credit improves further.}  The local reward (Section~\ref{sec:method}) pays for accepted suppression on or next to burning cells, including wetline, without replay. It reaches $24.7\%$ and beats the shared reward in all eight seed pairs. Simulator credit ($20.9\%$) beats the local reward in all eight pairs, by  3.9 points on average ($95\%$ interval $[2.1, 5.7]$ from a paired $t$ test; Appendix~\ref{app:local-control}).  The local reward pays for each overlapping placement, and for suppression near the fire even when it saves nothing in that step. Simulator credit pays only for the burn that a placement prevents.

\textbf{The gain remains under two rule changes.} We retrained with eight seeds after making ground crews carry and refill their water, and after letting firelines also start from burned ground. Simulator credit beats the shared reward in all eight seed pairs under each change. With both changes, it scores $11.3\%$, against $21.2\%$ for the shared reward and $17.5\%$ for the local reward, and wins all eight pairs against both (Appendix~\ref{app:added-experiments}).

Learned credit, using informed critics, scores $27.0\%$ and loses all eight pairs to simulator credit. Its mean difference from shared reward is  0.40 percentage points ($90\%$ interval $[0.12, 0.69]$ from a paired $t$ test), inside the  equivalence margin of $\pm1$ point fixed before inspecting these outcomes. Both one-sided tests (TOST) reject at $\alpha=.05$ \citep{schuirmann1987equivalence}.  So learned credit does no better than the shared reward, which keeps a small estimated lead. We do not call this equivalence, because half of the actors trained with learned credit were saturated.

Four of the eight  actors trained with learned credit (seeds 2, 3, 5 and 6) are saturated on  every trajectory of their final policies that we checked: every action score lies outside the clamp, so the policy no longer changes (Section~\ref{sec:method}). Seeds 2 and 3 end with the same fixed action and therefore score identically on all thirty fires.  Three repairs leave learned credit no better than the shared reward. Without the clamp, under the current rules, saturation cannot occur and most of the retrained actors keep learning, yet all eight burn more than the shared reward retrained alongside them with the same seeds ($27.3\%$ against $26.6\%$). Training the critic also on replayed outcomes of untaken actions over one step gives $26.9\%$, again worse than that shared reward in all eight pairs, and a separate output head for each agent type does not lower the burn score in a trial with three seeds (Appendix~\ref{app:added-experiments}).  

\begin{figure}[htbp]
\centering
\includegraphics[width=0.8\linewidth]{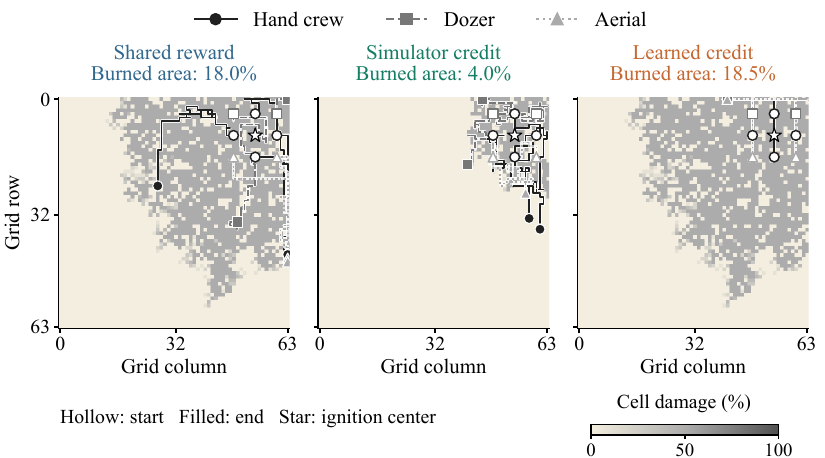}
\caption[One fire chosen in advance.]{ One fire chosen in advance (seed 1, first evaluation fire): damaged cells (gray) and agent paths. With simulator credit, crews stay at the fire's edge and the fire burns out at step 69 (burn score $4.0\%$); with the shared reward and with learned credit, it spreads until step 100 (burn score about $18\%$).}
\label{fig:fire-paths}
\end{figure}

 A comparison under earlier rules also favors simulator credit in all eight seed pairs and over the two other trained arms across five simulated scenarios (Appendix~\ref{app:performance}). On the preselected fire in Figure~\ref{fig:fire-paths},  ground crews trained with simulator credit favor immediate wetline attack.

\begin{figure}[!t]
\centering
\includegraphics[width=\linewidth]{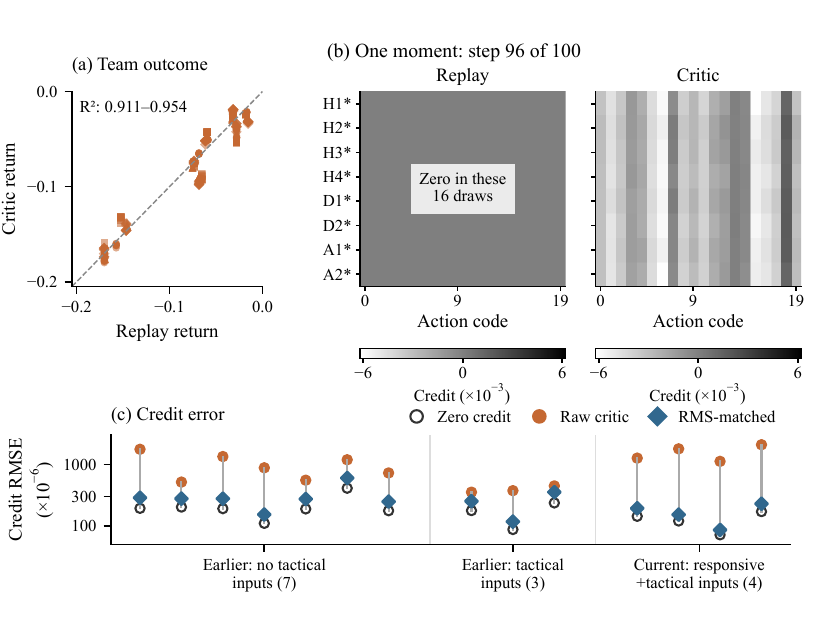}
\caption[Return fit and credit calibration.]{Return fit and credit calibration.  (a) Predicted against replayed return for the actions taken, for the four critics trained under the current rules. (b) A moment at which all actions give the same return in all sixteen replays, so the replay credit is zero (stars), while the critic's credit reaches $6.1\times10^{-3}$.  (c) Error of the raw critic credit (orange), of the critic credit rescaled to the size of the replay credit (blue) and of zero credit (open circles) for fourteen checkpoints; the raw error exceeds zero's at every checkpoint (Appendices~\ref{app:completed-controls}, \ref{app:earlier-controls} and~\ref{app:scale-sensitivity}).}
\label{fig:informed-audit}
\end{figure}

\begin{samepage}
\subsection{Does learned credit match the replay?}
\label{sec:audit-results}
\textbf{Critics trained under the current rules track returns but miscalibrate credit.} We audit four informed critics paired with non-saturated actors from Section~\ref{sec:learning-results} (seeds 1, 4, 7, 8) under current rules; none of their 256 audited decisions is fully saturated.  They predict the team's return well ($R^2$ from $0.91$ to $0.95$), yet the credit they assign is 9 to 16 times too large: its RMSE is $8.8$ to $15.6$ times that of giving every action zero credit, whose error equals the size of the replay credit (Figure~\ref{fig:informed-audit}; Tables~\ref{tab:current-audit-calibration} and~\ref{tab:current-audit-difference}). The $95\%$ interval for $\Delta$ is above zero for all four critics. At moments earlier in the episode, raising the replay budget from 16 to 64 replays per action raises this ratio to between 20 and 42 as replay noise falls. Ten critics trained under earlier rules show the same pattern, with $R^2$ up to $0.97$ (Appendix~\ref{app:earlier-controls}).
\par
\end{samepage}

\textbf{Critic rankings are highly similar across agents.}  The median Spearman correlation between agents' rankings is $0.98$ to $0.99$, so each critic ranks the 20 actions almost identically for hand crews, dozers and aircraft. Untrained critics of the same design already agree as closely ($0.98$ to $0.99$), so training does not make the ranking specific to each agent. Separately, 19 to 47 of each critic's 64 cases (eight agents at eight moments) have equal returns for all 20 actions in every sampled replay, yet the critic still assigns them credit (Figure~\ref{fig:informed-audit}b).

\textbf{Precision of the replay reference.}  Seven of the eight audited moments fall late in the episode ($t\ge55$), where the replay credit is about as large as its own noise with sixteen replays (Table~\ref{tab:current-audit-calibration}). With 64 replays, earlier moments are only slightly cleaner (signal-to-noise ratio 1.1 to 1.4, against 0.9 to 1.2 at the original eight), and for one critic even 256 replays per action fall short of the precision threshold we set in advance: the lower end of the $95\%$ interval of the signal-to-noise ratio had to reach $\sqrt{2}$, where signal equals noise, since the mean square of the replay credit adds the squared Monte Carlo error to the squared signal. The order of each agent's actions, on which the rescaled comparisons also depend, therefore stays unresolved (Appendices~\ref{app:audit}, \ref{app:scale-sensitivity} and~\ref{app:added-experiments}).

\subsection{Ablations and controls}
\label{sec:ablations}
\textbf{Common random numbers reduce credit noise.}  In a test under earlier rules on 32 saved moments, giving the two replays of a step independent fire draws makes the variance of simulator credit $934$ times larger (Figure~\ref{fig:repaired-learning}c). With independent draws, $46.9\%$ of the credit samples are below zero, against none with shared draws (Appendix~\ref{app:offline}).

\begin{samepage}
 Replacing the agent's whole action, by the same move without suppression or by staying in place, gives exactly the same simulator credit as deleting only its suppression in all 256 sampled cases (Appendix~\ref{app:offline}). None of the twelve critic settings we tried under earlier rules beats the shared reward (Appendix~\ref{app:supplement}).  
\par
\end{samepage}

\section{Discussion and conclusion}
\label{sec:discussion}
Simulator credit reduces MAPPO's IQM burn score from $26.6\%$ with the shared reward and $24.7\%$ with a local reward to $20.9\%$, winning every paired seed against both, and the gain remains under two rule changes. Learned credit does no better than the shared reward.  Four trained critics predict the team's return well ($R^2$ above $0.9$), yet the credit they assign is 9 to 42 times larger than the replay credit, depending on the replay budget, and they rank actions almost identically for every agent type.  Predicting returns well is therefore a weak test of a critic's credit.

\paragraph{Scope and future directions.}
 These findings concern MAPPO trained with each feedback signal and the COMA-style critics we trained in FLARE. Resource limits other than the water limit of Section~\ref{sec:learning-results}, and other forms of local reward, remain untested, and grading the order of an agent's actions will need larger replay budgets or estimates with less noise. Because FLARE can replay any moment to the end of the episode, simulator credit can also be computed over several steps, which would credit a fireline for the spread it blocks after the step in which it is placed. Critics trained on replayed outcomes over more than one step also remain to be tested.

\clearpage
\section*{Reproducibility statement}
Section~\ref{sec:benchmark} defines the environment and Table~\ref{tab:experiment-inventory} maps each reported study to its rules and sample counts. The main comparison archives source/configuration records, complete logs of every training restart, checkpoints and fixed fresh-fire evaluations. Earlier comparisons retain actor checkpoints, evaluation arrays and training summaries; logs and per-run summaries document the auxiliary controls. The online audits reuse the actors and critics of four main-comparison runs whose actors were not saturated and of ten earlier-rule runs, with eight moments and sixteen replays per action. They archive checkpoints, saved normalizers and replay arrays; scale and frozen-gradient diagnostics additionally bind cross-validation, resampling and gradient outputs. Each figure's source records the hashes of the saved arrays and logs it uses. Sections~\ref{sec:method} and~\ref{sec:protocol} define learning and aggregation; Appendices~\ref{app:config} and~\ref{app:audit} provide settings, sampling units and uncertainty calculations.

\section*{Ethics statement}
The simulator is an abstract research environment. Its policies have not been validated for deployment or incident command, and the results should not guide real suppression operations.

\section*{AI use statement}
Generative AI tools assisted with literature retrieval, code inspection and implementation, experiment and diagnostic design, interpretation of recorded results, formulation and checking of mathematical arguments, figure-generation code, and manuscript drafting and editing. Numerical figures in this paper are computed from recorded experiments. The use of automated checks and source comparisons does not replace the authors' responsibility for the final claims and submission.

\bibliography{iclr2027_submission}

\begin{thebibliography}{36}
\providecommand{\natexlab}[1]{#1}
\providecommand{\url}[1]{\texttt{#1}}
\expandafter\ifx\csname urlstyle\endcsname\relax
  \providecommand{\doi}[1]{doi: #1}\else
  \providecommand{\doi}{doi: \begingroup \urlstyle{rm}\Url}\fi

\bibitem[Agarwal et~al.(2021)Agarwal, Schwarzer, Castro, Courville, and
  Bellemare]{agarwal2021precipice}
Rishabh Agarwal, Max Schwarzer, Pablo~Samuel Castro, Aaron Courville, and
  Marc~G. Bellemare.
\newblock Deep reinforcement learning at the edge of the statistical precipice.
\newblock In \emph{Advances in Neural Information Processing Systems}, 2021.
\newblock URL \url{https://arxiv.org/abs/2108.13264}.

\bibitem[Buesing et~al.(2019)Buesing, Weber, Zwols, Heess, Racaniere, Guez, and
  Lespiau]{buesing2019woulda}
Lars Buesing, Theophane Weber, Yori Zwols, Nicolas Heess, Sebastien Racaniere,
  Arthur Guez, and Jean-Baptiste Lespiau.
\newblock Woulda, coulda, shoulda: Counterfactually-guided policy search.
\newblock In \emph{International Conference on Learning Representations}, 2019.
\newblock URL \url{https://openreview.net/forum?id=BJG0voC9YQ}.

\bibitem[{\c C}ak{\i}r et~al.(2025){\c C}ak{\i}r, Darvariu, Lacerda, and
  Hawes]{cakir2025jaxwildfire}
Ufuk {\c C}ak{\i}r, Victor-Alexandru Darvariu, Bruno Lacerda, and Nick Hawes.
\newblock {JaxWildfire}: A {GPU}-accelerated wildfire simulator for
  reinforcement learning.
\newblock \emph{arXiv preprint arXiv:2512.06102}, 2025.
\newblock URL \url{https://arxiv.org/abs/2512.06102}.

\bibitem[Castellini et~al.(2025)Castellini, Devlin, Oliehoek, and
  Savani]{castellini2022difference}
Jacopo Castellini, Sam Devlin, Frans~A. Oliehoek, and Rahul Savani.
\newblock Difference rewards policy gradients.
\newblock \emph{Neural Computing and Applications}, 37:\penalty0 13163--13186,
  2025.
\newblock \doi{10.1007/s00521-022-07960-5}.
\newblock URL
  \url{https://link.springer.com/article/10.1007/s00521-022-07960-5}.
\newblock First published online in 2022.

\bibitem[Chen et~al.(2026)Chen, Sun, Wang, Wang, Zhang, Shen, Li, and
  Zhang]{chen2026exact}
Yanjun Chen, Yirong Sun, Hanlin Wang, Jinghan Wang, Xinming Zhang, Xiaoyu Shen,
  Wenjie Li, and Wei Zhang.
\newblock Exact is easier: Credit assignment for cooperative {LLM} agents,
  2026.
\newblock URL \url{https://arxiv.org/abs/2603.06859v2}.

\bibitem[Colby et~al.(2015)Colby, Kharaghani, HolmesParker, and
  Tumer]{colby2015counterfactual}
Mitchell~K. Colby, Sepideh Kharaghani, Chris HolmesParker, and Kagan Tumer.
\newblock Counterfactual exploration for improving multiagent learning.
\newblock In \emph{Proceedings of the 2015 International Conference on
  Autonomous Agents and Multiagent Systems (AAMAS)}, pp.\  171--179, Istanbul,
  Turkey, 2015. IFAAMAS.

\bibitem[Deshmukh et~al.(2026)Deshmukh, Subramanian, Addanki, and
  Vlassis]{deshmukh2026cosac}
Shripad Deshmukh, Jayakumar Subramanian, Raghavendra Addanki, and Nikos
  Vlassis.
\newblock {COSAC}: Counterfactual credit assignment in sequential cooperative
  teams, 2026.
\newblock URL \url{https://arxiv.org/abs/2604.17693v2}.

\bibitem[Foerster et~al.(2018)Foerster, Farquhar, Afouras, Nardelli, and
  Whiteson]{foerster2018coma}
Jakob Foerster, Gregory Farquhar, Triantafyllos Afouras, Nantas Nardelli, and
  Shimon Whiteson.
\newblock Counterfactual multi-agent policy gradients.
\newblock In \emph{Proceedings of the AAAI Conference on Artificial
  Intelligence}, volume~32, 2018.
\newblock \doi{10.1609/aaai.v32i1.11794}.
\newblock URL \url{https://ojs.aaai.org/index.php/AAAI/article/view/11794}.

\bibitem[Fujimoto et~al.(2019)Fujimoto, Meger, and Precup]{fujimoto2019bcq}
Scott Fujimoto, David Meger, and Doina Precup.
\newblock Off-policy deep reinforcement learning without exploration.
\newblock In \emph{Proceedings of the 36th International Conference on Machine
  Learning}, volume~97 of \emph{Proceedings of Machine Learning Research}, pp.\
   2052--2062, 2019.
\newblock URL \url{https://proceedings.mlr.press/v97/fujimoto19a.html}.

\bibitem[Glasserman \& Yao(1992)Glasserman and Yao]{glasserman1992common}
Paul Glasserman and David~D. Yao.
\newblock Some guidelines and guarantees for common random numbers.
\newblock \emph{Management Science}, 38\penalty0 (6):\penalty0 884--908, 1992.
\newblock \doi{10.1287/mnsc.38.6.884}.
\newblock URL \url{https://pubsonline.informs.org/doi/10.1287/mnsc.38.6.884}.

\bibitem[Haksar \& Schwager(2018)Haksar and Schwager]{haksar2018wildfire}
Ravi~N. Haksar and Mac Schwager.
\newblock Distributed deep reinforcement learning for fighting forest fires
  with a network of aerial robots.
\newblock In \emph{2018 IEEE/RSJ International Conference on Intelligent Robots
  and Systems (IROS)}, pp.\  1067--1074. IEEE, 2018.
\newblock \doi{10.1109/IROS.2018.8593539}.
\newblock URL
  \url{https://msl.stanford.edu/bibliography/haksar_distributed_2018}.

\bibitem[HolmesParker et~al.(2014)HolmesParker, Taylor, Agogino, and
  Tumer]{holmesparker2014clean}
Chris HolmesParker, Matthew~E. Taylor, Adrian~K. Agogino, and Kagan Tumer.
\newblock {CLEAN} rewards to improve coordination by removing exploratory
  action noise.
\newblock In \emph{2014 IEEE/WIC/ACM International Joint Conferences on Web
  Intelligence (WI) and Intelligent Agent Technologies (IAT)}, pp.\  127--134,
  Warsaw, Poland, 2014. IEEE.
\newblock \doi{10.1109/WI-IAT.2014.159}.

\bibitem[Hyun et~al.(2025)Hyun, Waytowich, and Chen]{hyun2025crew}
Jonathan Hyun, Nicholas~R. Waytowich, and Boyuan Chen.
\newblock {CREW-Wildfire}: Benchmarking agentic multi-agent collaborations at
  scale.
\newblock \emph{Transactions on Machine Learning Research}, 2025.
\newblock URL \url{https://openreview.net/forum?id=8mr27qFzKR}.

\bibitem[Ilyas et~al.(2020)Ilyas, Engstrom, Santurkar, Tsipras, Janoos,
  Rudolph, and Madry]{ilyas2020closer}
Andrew Ilyas, Logan Engstrom, Shibani Santurkar, Dimitris Tsipras, Firdaus
  Janoos, Larry Rudolph, and Aleksander Madry.
\newblock A closer look at deep policy gradients.
\newblock In \emph{International Conference on Learning Representations}, 2020.
\newblock URL \url{https://openreview.net/forum?id=ryxdEkHtPS}.

\bibitem[Kuba et~al.(2021)Kuba, Wen, Meng, Gu, Zhang, Mguni, Wang, and
  Yang]{kuba2021variance}
Jakub~Grudzien Kuba, Muning Wen, Linghui Meng, Shangding Gu, Haifeng Zhang,
  David~Henry Mguni, Jun Wang, and Yaodong Yang.
\newblock Settling the variance of multi-agent policy gradients.
\newblock In \emph{Advances in Neural Information Processing Systems},
  volume~34, pp.\  13458--13470, 2021.
\newblock URL
  \url{https://proceedings.neurips.cc/paper/2021/hash/6fe6a8a6e6cb710584efc4af0c34ce50-Abstract.html}.

\bibitem[Li et~al.(2026)Li, Tian, Chen, Zhang, Liu, Ban, and
  Zhuang]{li2026ccpo}
Zhongyi Li, Wan Tian, Jinju Chen, Huiming Zhang, Yang Liu, Yikun Ban, and
  Fuzhen Zhuang.
\newblock Counterfactual credit policy optimization for multi-agent
  collaboration, 2026.
\newblock URL \url{https://arxiv.org/abs/2603.21563v5}.

\bibitem[Lyu et~al.(2023)Lyu, Baisero, Xiao, Daley, and
  Amato]{lyu2023centralized}
Xueguang Lyu, Andrea Baisero, Yuchen Xiao, Brett Daley, and Christopher Amato.
\newblock On centralized critics in multi-agent reinforcement learning.
\newblock \emph{Journal of Artificial Intelligence Research}, 77:\penalty0
  295--354, 2023.
\newblock \doi{10.1613/jair.1.14386}.
\newblock URL \url{https://www.jair.org/index.php/jair/article/view/14386}.

\bibitem[Mesnard et~al.(2021)Mesnard, Weber, Viola, Thakoor, Saade,
  Harutyunyan, Dabney, Stepleton, Heess, Guez, Moulines, Hutter, Buesing, and
  Munos]{mesnard2021counterfactual}
Thomas Mesnard, Theophane Weber, Fabio Viola, Shantanu Thakoor, Alaa Saade,
  Anna Harutyunyan, Will Dabney, Thomas~S. Stepleton, Nicolas Heess, Arthur
  Guez, Eric Moulines, Marcus Hutter, Lars Buesing, and Remi Munos.
\newblock Counterfactual credit assignment in model-free reinforcement
  learning.
\newblock In \emph{Proceedings of the 38th International Conference on Machine
  Learning}, volume 139 of \emph{Proceedings of Machine Learning Research},
  pp.\  7654--7664, 2021.
\newblock URL \url{https://proceedings.mlr.press/v139/mesnard21a.html}.

\bibitem[Oberst \& Sontag(2019)Oberst and Sontag]{oberst2019counterfactual}
Michael Oberst and David Sontag.
\newblock Counterfactual off-policy evaluation with {Gumbel-Max} structural
  causal models.
\newblock In \emph{Proceedings of the 36th International Conference on Machine
  Learning}, volume~97 of \emph{Proceedings of Machine Learning Research}, pp.\
   4881--4890. PMLR, 2019.
\newblock URL \url{https://proceedings.mlr.press/v97/oberst19a.html}.

\bibitem[Papoudakis et~al.(2021)Papoudakis, Christianos, Sch{\"a}fer, and
  Albrecht]{papoudakis2021benchmarking}
Georgios Papoudakis, Filippos Christianos, Lukas Sch{\"a}fer, and Stefano~V.
  Albrecht.
\newblock Benchmarking multi-agent deep reinforcement learning algorithms in
  cooperative tasks.
\newblock In \emph{Proceedings of the Neural Information Processing Systems
  Track on Datasets and Benchmarks}, volume~1, 2021.
\newblock URL
  \url{https://datasets-benchmarks-proceedings.neurips.cc/paper_files/paper/2021/hash/a8baa56554f96369ab93e4f3bb068c22-Abstract-round1.html}.

\bibitem[Patino et~al.(2025)Patino, Billings, Saleh~Abadi, Redder, Eck, Doshi,
  and Soh]{patino2025moasei}
Ceferino Patino, Tyler~J. Billings, Alireza Saleh~Abadi, Daniel Redder, Adam
  Eck, Prashant Doshi, and Leen-Kiat Soh.
\newblock Inaugural {MOASEI} competition at {AAMAS'2025}: A technical report.
\newblock \emph{arXiv preprint arXiv:2507.05469}, 2025.
\newblock URL \url{https://arxiv.org/abs/2507.05469}.

\bibitem[Rahmattalabi et~al.(2016)Rahmattalabi, Chung, Colby, and
  Tumer]{rahmattalabi2016dpp}
Aida Rahmattalabi, Jen~Jen Chung, Mitchell~K. Colby, and Kagan Tumer.
\newblock {D++}: Structural credit assignment in tightly coupled multiagent
  domains.
\newblock In \emph{2016 {IEEE/RSJ} International Conference on Intelligent
  Robots and Systems ({IROS})}, pp.\  4424--4429, Daejeon, South Korea, 2016.
  IEEE.
\newblock \doi{10.1109/IROS.2016.7759651}.

\bibitem[Rashid et~al.(2018)Rashid, Samvelyan, Schroeder~de Witt, Farquhar,
  Foerster, and Whiteson]{rashid2018qmix}
Tabish Rashid, Mikayel Samvelyan, Christian Schroeder~de Witt, Gregory
  Farquhar, Jakob Foerster, and Shimon Whiteson.
\newblock {QMIX}: Monotonic value function factorisation for deep multi-agent
  reinforcement learning.
\newblock In \emph{Proceedings of the 35th International Conference on Machine
  Learning}, volume~80 of \emph{Proceedings of Machine Learning Research}, pp.\
   4295--4304, 2018.
\newblock URL \url{https://proceedings.mlr.press/v80/rashid18a.html}.

\bibitem[Schuirmann(1987)]{schuirmann1987equivalence}
Donald~J. Schuirmann.
\newblock A comparison of the two one-sided tests procedure and the power
  approach for assessing the equivalence of average bioavailability.
\newblock \emph{Journal of Pharmacokinetics and Biopharmaceutics}, 15:\penalty0
  657--680, 1987.
\newblock \doi{10.1007/BF01068419}.
\newblock URL \url{https://link.springer.com/article/10.1007/BF01068419}.

\bibitem[Schulman et~al.(2015)Schulman, Levine, Abbeel, Jordan, and
  Moritz]{schulman2015trpo}
John Schulman, Sergey Levine, Pieter Abbeel, Michael Jordan, and Philipp
  Moritz.
\newblock Trust region policy optimization.
\newblock In \emph{Proceedings of the 32nd International Conference on Machine
  Learning}, volume~37 of \emph{Proceedings of Machine Learning Research}, pp.\
   1889--1897, 2015.

\bibitem[Schulman et~al.(2016)Schulman, Moritz, Levine, Jordan, and
  Abbeel]{schulman2016gae}
John Schulman, Philipp Moritz, Sergey Levine, Michael~I. Jordan, and Pieter
  Abbeel.
\newblock High-dimensional continuous control using generalized advantage
  estimation.
\newblock In \emph{International Conference on Learning Representations}, 2016.
\newblock URL \url{https://arxiv.org/abs/1506.02438}.

\bibitem[Schulman et~al.(2017)Schulman, Wolski, Dhariwal, Radford, and
  Klimov]{schulman2017proximal}
John Schulman, Filip Wolski, Prafulla Dhariwal, Alec Radford, and Oleg Klimov.
\newblock Proximal policy optimization algorithms.
\newblock \emph{arXiv preprint arXiv:1707.06347}, 2017.
\newblock URL \url{https://arxiv.org/abs/1707.06347}.

\bibitem[Sunehag et~al.(2018)Sunehag, Lever, Gruslys, Czarnecki, Zambaldi,
  Jaderberg, Lanctot, Sonnerat, Leibo, Tuyls, and Graepel]{sunehag2018vdn}
Peter Sunehag, Guy Lever, Audrunas Gruslys, Wojciech~Marian Czarnecki, Vinicius
  Zambaldi, Max Jaderberg, Marc Lanctot, Nicolas Sonnerat, Joel~Z. Leibo, Karl
  Tuyls, and Thore Graepel.
\newblock Value-decomposition networks for cooperative multi-agent learning
  based on team reward.
\newblock In \emph{Proceedings of the 17th International Conference on
  Autonomous Agents and MultiAgent Systems}, pp.\  2085--2087. International
  Foundation for Autonomous Agents and Multiagent Systems, 2018.
\newblock URL
  \url{https://www.ifaamas.org/Proceedings/aamas2018/pdfs/p2085.pdf}.
\newblock Extended abstract.

\bibitem[Tumer \& Agogino(2007)Tumer and Agogino]{tumer2007distributed}
Kagan Tumer and Adrian Agogino.
\newblock Distributed agent-based air traffic flow management.
\newblock In \emph{Proceedings of the 6th International Joint Conference on
  Autonomous Agents and Multiagent Systems (AAMAS)}, pp.\  330--337, Honolulu,
  HI, USA, 2007. IFAAMAS.
\newblock \doi{10.1145/1329125.1329434}.

\bibitem[van Hasselt et~al.(2016)van Hasselt, Guez, Hessel, Mnih, and
  Silver]{vanhasselt2016popart}
Hado van Hasselt, Arthur Guez, Matteo Hessel, Volodymyr Mnih, and David Silver.
\newblock Learning values across many orders of magnitude.
\newblock In \emph{Advances in Neural Information Processing Systems}, 2016.
\newblock URL \url{https://arxiv.org/abs/1602.07714}.

\bibitem[Wang et~al.(2021{\natexlab{a}})Wang, Ren, Liu, Yu, and
  Zhang]{wang2021qplex}
Jianhao Wang, Zhizhou Ren, Terry Liu, Yang Yu, and Chongjie Zhang.
\newblock {QPLEX}: Duplex dueling multi-agent {Q}-learning.
\newblock In \emph{International Conference on Learning Representations},
  2021{\natexlab{a}}.
\newblock URL \url{https://openreview.net/forum?id=Rcmk0xxIQV}.

\bibitem[Wang et~al.(2020)Wang, Zhang, Kim, and Gu]{wang2020shapley}
Jianhong Wang, Yuan Zhang, Tae-Kyun Kim, and Yunjie Gu.
\newblock {Shapley Q-Value}: A local reward approach to solve global reward
  games.
\newblock In \emph{Proceedings of the AAAI Conference on Artificial
  Intelligence}, volume~34, pp.\  7285--7292, 2020.
\newblock \doi{10.1609/aaai.v34i05.6220}.
\newblock URL \url{https://ojs.aaai.org/index.php/AAAI/article/view/6220}.

\bibitem[Wang et~al.(2021{\natexlab{b}})Wang, Han, Wang, Dong, and
  Zhang]{wang2021dop}
Yihan Wang, Beining Han, Tonghan Wang, Heng Dong, and Chongjie Zhang.
\newblock {DOP}: Off-policy multi-agent decomposed policy gradients.
\newblock In \emph{International Conference on Learning Representations},
  2021{\natexlab{b}}.
\newblock URL \url{https://openreview.net/forum?id=6FqKiVAdI3Y}.

\bibitem[Wolpert \& Tumer(1999)Wolpert and Tumer]{wolpert1999introduction}
David~H. Wolpert and Kagan Tumer.
\newblock An introduction to collective intelligence.
\newblock \emph{arXiv preprint cs/9908014}, 1999.
\newblock URL \url{https://arxiv.org/abs/cs/9908014}.

\bibitem[Yu et~al.(2022)Yu, Velu, Vinitsky, Gao, Wang, Bayen, and
  Wu]{yu2022surprising}
Chao Yu, Akash Velu, Eugene Vinitsky, Jiaxuan Gao, Yu~Wang, Alexandre Bayen,
  and Yi~Wu.
\newblock The surprising effectiveness of {PPO} in cooperative multi-agent
  games.
\newblock In \emph{Advances in Neural Information Processing Systems},
  volume~35, 2022.
\newblock URL
  \url{https://proceedings.neurips.cc/paper_files/paper/2022/hash/9c1535a02f0ce079433344e14d910597-Abstract.html}.

\bibitem[Zhao \& Xie(2025)Zhao and Xie]{zhao2025maca}
Xutong Zhao and Yaqi Xie.
\newblock Multi-level advantage credit assignment for cooperative multi-agent
  reinforcement learning.
\newblock In \emph{Proceedings of The 28th International Conference on
  Artificial Intelligence and Statistics}, volume 258 of \emph{Proceedings of
  Machine Learning Research}, pp.\  2971--2979, 2025.
\newblock URL \url{https://proceedings.mlr.press/v258/zhao25c.html}.

\end{thebibliography}
\bibliographystyle{iclr2027_conference}

\clearpage
\appendix
\renewcommand{\topfraction}{.95}
\renewcommand{\bottomfraction}{.95}
\renewcommand{\textfraction}{.05}
\renewcommand{\floatpagefraction}{.8}
\setcounter{topnumber}{8}
\setcounter{bottomnumber}{8}
\setcounter{totalnumber}{12}

\section{Environment, training and evaluation configuration}
\label{app:config}
\suppressfloats[t]

\begin{table}[!htbp]
\centering
\begin{tabular}{ll}
\toprule
Setting & Value \\
\midrule
Training scenario /  seeds per feedback signal & S1 / 8 \\
Grid / roster / maximum episode steps & $64\times64$ / $4$ hand crew, $2$ dozer, $2$  aircraft / $100$ \\
Parallel environments / rollout length & $4$ / $128$ \\
Agent-steps per update / number of updates & $4096$ / $732$ \\
Actor and base critic hidden width & $128$ \\
Actor / critic learning rate & $5\times10^{-4}$ / $5\times10^{-4}$ \\
Discount / GAE or TD($\lambda$) parameter $\lambda$ & $0.99$ / $0.95$ \\
PPO clip / entropy coefficient & $0.2$ / $0.01$ \\
Actor logits in the main runs & Hard clamp to $[-10,10]$ \\
Minibatch / maximum PPO epochs & $64$ / $4$ \\
KL early-stop target / gradient clipping & $0.01$ / $10$ \\
Reward normalization & Running standard deviation \\
Evaluation / fires per policy &  Most likely action / $30$ \\
Reported metric &  Final burn score \\
\bottomrule
\end{tabular}
\caption[Training and evaluation settings.]{Training and evaluation settings for the heterogeneous team;  the runs without the clamp change only the bound on the actor's logits. Actor settings are shared across arms; V and Q critics and their targets differ as described in the main text.}
\end{table}

\begin{table}[!htbp]
\centering
\small
\begin{tabular}{p{0.36\linewidth}p{0.57\linewidth}}
\toprule
Environment setting & Main heterogeneous configuration (S1) \\
\midrule
Grid / episode limit & $64\times64$ cells / 100 steps \\
Roster & Four hand crews, two dozers,  two aircraft \\
Initial ignition & One randomly located $5\times5$ square; radius parameter $2$ \\
Initial fuel / slope & Fuel $1$ per cell / slope $0$ \\
Agent placement & On a ring six cells from the ignition \\
Local actor input & Own-centered $21\times21$ burning, mitigation, and fuel crops \\
Other actor inputs & Coarse burning mean and uncertainty; own type, water fraction, refill flag, global no-fly flag \\
Ensemble Kalman filter & Disabled; belief mean equals simulated state and uncertainty is zero \\
Action labels & Five movements $\times$ four suppression primitives $=20$ \\
Ground / aerial constraints & Enabled / enabled \\
Aerial water capacity / debit & $8$ units / $1$ per accepted gated placement \\
Refill cooldown & $5$ blocked steps including the empty-tank request that triggers refill \\
No-fly onset probability & $0.1$, sampled only if the timer is inactive at step entry \\
No-fly duration & Uniform integer in $\{3,4,5,6\}$; onset step counts toward duration \\
Ground--aerial separation & Chebyshev radius $1$ around the aerial target center \\
Aerial primary footprint & Patch radius $1$; up to $3\times3$ distinct cells \\
Dozer primary footprint & Line length $3$ in movement direction; staying affects one cell \\
Wet mitigation & Strength $0.6$ for one fire transition; cleared afterwards \\
Scratchline effect & Multiply remaining fuel by $0.5$ \\
Ground fireline effect & Mitigation strength $1$; multiply remaining fuel by $0$ \\
Mitigation mode / reward shaping & Treatment effect applied after the spread probability is computed / additional shaping disabled \\
Wind in spread calculation & Magnitude only; no directional transport \\
Aerial suppression checks & Apply to every suppression primitive; one water unit per accepted placement \\
\bottomrule
\end{tabular}
\caption[Environment settings of the main scenario.]{Settings of the main scenario (S1); Appendix~\ref{app:performance} states the other scenarios. Action semantics are defined in Section~\ref{sec:benchmark}. An empty-tank request reaching the water check starts refill; water returns after five blocked steps. Footprints are clipped at boundaries. Wet suppression clears pre-existing fireline markings on its footprint after the transition while preserving zero fuel, so neighboring ground targets can lose that source of anchoring.}
\label{tab:environment-settings}
\end{table}

\subsection{Transition and replay}
\label{app:transition-details}
The simulator  updates the time and no-fly state, moves agents, accepts placements, combines their effects and advances fire. Combination uses maximum mitigation, minimum fuel multipliers and the union of wet footprints;  after the fire spreads, wet mitigation clears and refill timers count down.  A ground primary fireline needs the map edge or an eight-neighbor that held a fireline before the transition, so a same-step placement cannot anchor another crew. Ground wetline and scratchline are unanchored; wetline has no modeled water cost and extinguishes its cell during the transition.  An aerial placement is accepted only if neither a no-fly period nor a refill is active, water remains and no ground unit stands within Chebyshev distance one of the target after movement.  Replacing a whole action in replay checks acceptance again, while deleting a placement keeps acceptance fixed.  Restoring a saved moment also restores the tactical state and the states of all random number generators. Given a frozen policy and  the underlying random draws of the environment and the policy, each replay is deterministic,  with observations specific to each branch and zero reward after the episode ends.

\subsection{Experiment inventory and rule sets}
\label{app:experiment-inventory}
Current rules apply all aerial checks and water debits to every suppression primitive. Earlier rules  applied them only to the main patch drop; aerial one-cell wetline and scratchline were exempt.  Ground rules are the same under both rule sets. Every evaluation and audit matches its policy's training rules. Training seeds identify policies; episodes, actions and replays are repeated measurements within them.
\begin{table}[!htbp]
\centering\small
\setlength{\tabcolsep}{4pt}
\begin{tabular}{@{}p{.27\linewidth}p{.12\linewidth}p{.56\linewidth}@{}}
\toprule
Study & Rules & Purpose and sample units \\
\midrule
Main learning + local control & Current & 8 seeds per 4 methods, 30 fresh fires each; local CPU, other arms CUDA. Profiles reuse evaluations; paths/action mix reuse seed 1 on the first fire. \\
Scenario transfer & Earlier & 24 policies (8 seeds per 3 methods), 5 scenarios, 30 fires each; S1 reused. Heuristic/no-suppression references: 30 fires per scenario. \\
Responsive main audit & Current & Reuse learned-credit seeds 1, 4, 7, 8; 8 contexts, 16 replays per action per critic. \\
Consistency / information & Earlier & 10 trained critic/actor pairs: 3 base,  1 tuned, 3 informed, 3 capacity-matched zero-input; 8 contexts, 16 replays per action each. \\
Audit sensitivity & Both & Reuse all 14 audits for scale, weighting and reference checks; frozen gradients use the current 4. Checkpoints analyzed separately. \\
Critic tuning & Earlier & 12 configurations $\times$ 3 seeds;  means over the last ten training updates. \\
No-clamp training & Earlier &  6 runs paired with the 3 informed and 3 zero-input runs; frozen-policy evaluation. \\
Randomness / removal & Earlier & Reuse 8 simulator actors, 4 contexts each; 256 paired draw sets per context; 2 action-replacement baselines on 256 cases, one per agent and context. \\
Actor saturation & Both & Reuse 16 earlier MAPPO, 8 current learned-credit and 8 local actors; 4 trajectories and selected gradient contexts per actor. \\
\bottomrule
\end{tabular}
\caption[Reported experiments and policy reuse.]{Reported experiments and policy reuse. Full training runs of the heterogeneous team complete 732 updates and 2,998,272 agent-steps. A context is a saved moment of an episode: the complete simulator state at one step, including the random number generator states.}
\label{tab:experiment-inventory}
\end{table}

\section{Performance and scenario transfer}
\label{app:performance}
 All policies trained under the current rules take their most likely action on fires 2026092400--2026092429,  which the saved logs of training resets show were never used in training. Each seed's score averages thirty fires;  the interquartile mean (IQM) averages the middle four of the eight seed scores. Unless stated otherwise, $95\%$ intervals  come from 10,000 bootstrap resamples of training seeds, paired for comparisons and conditional on these fixed fires.
\begin{table}[!htbp]
\centering
\small
\setlength{\tabcolsep}{4pt}
\renewcommand{\arraystretch}{1.12}
\begin{tabular}{lrr}
\toprule
Method &  Burn score (IQM) & Descriptive $95\%$ interval \\
\midrule
Shared reward (MAPPO) & $0.2661$ & $[0.2613,0.2691]$ \\
Simulator credit (MAPPO) & $0.2085$ & $[0.1913,0.2280]$ \\
Learned credit & $0.2698$ & $[0.2662,0.2740]$ \\
Local reward & $0.2473$ & $[0.2400,0.2575]$ \\
\bottomrule
\end{tabular}
\caption{S1 performance: eight training seeds per arm, each evaluated on the same thirty fresh fires. Lower is better. Intervals resample training seeds and condition on the fixed fires. MAPPO feedback arms use shared, simulator or fixed local rewards; learned credit uses an informed Q critic. All arms retain the actor-logit clamp.}
\label{tab:repaired-headline}
\end{table}

\begin{table}[!htbp]
\centering
\small
\setlength{\tabcolsep}{4pt}
\renewcommand{\arraystretch}{1.12}
\begin{tabular}{rrrrrr}
\toprule
Seed & Shared & Simulator & Learned & Learned $-$ shared & Local \\
\midrule
1 & $0.271906$ & $0.189196$ & $0.266535$ & $-0.005371$ & $0.260199$ \\
2 & $0.267826$ & $0.212686$ & $0.275475$ & $+0.007650$ & $0.240091$ \\
3 & $0.269023$ & $0.174030$ & $0.275475$ & $+0.006452$ & $0.239967$ \\
4 & $0.261043$ & $0.198280$ & $0.266326$ & $+0.005282$ & $0.239664$ \\
5 & $0.261488$ & $0.253081$ & $0.263376$ & $+0.001888$ & $0.260044$ \\
6 & $0.260782$ & $0.203773$ & $0.268258$ & $+0.007476$ & $0.243898$ \\
7 & $0.266229$ & $0.226977$ & $0.271595$ & $+0.005366$ & $0.250273$ \\
8 & $0.268991$ & $0.219263$ & $0.272618$ & $+0.003627$ & $0.254977$ \\
\bottomrule
\end{tabular}
\caption[Burn score of each policy under the current rules.]{Mean burn score of each policy trained under the current rules, over the same thirty fires. Eight seed pairs are the replication units for method contrasts; together they aggregate the 240 episode outcomes per arm.}
\label{tab:repaired-seeds}
\end{table}

\paragraph[Prespecified equivalence test.]{Prespecified equivalence test.}
The $\pm0.01$  margin for learned credit minus shared reward was fixed before we inspected the outcomes under the current rules.  The paired $t$ test uses eight seed differences, with mean $0.004046$,  standard deviation $0.004265$,  standard error $0.001508$ and seven degrees of freedom. Its $90\%$ interval is $[0.001190,0.006903]$; one-sided $p$-values are $1.70\times10^{-5}$ and $0.00277$.  A paired bootstrap with 100,000 resamples, chosen in advance, gives $[0.001492,0.006122]$.
\paragraph{Fixed local-reward control.}
\label{app:local-control}
 The local reward, fixed in advance, counts the distinct cells of an agent's accepted placement within Chebyshev distance one of cells burning before the real transition, divided by $64^2$. It uses no counterfactual transition and adds no shared reward. The action set includes the same cost-free ground wetline, and accepted wetline cells near burning earn this local reward. Rejected placements receive zero; overlapping agents each receive their own footprint count.  Simulator credit instead measures the burning that a placement actually prevents, so redundant or immediately ineffective placements earn nothing.  The comparison therefore tests for a gain beyond this fixed reward for working near the fire, with the same access to wetline.

The eight local-reward runs use the same MAPPO hyperparameters, current rules, actor architecture and full training budget as the shared-reward and simulator-credit arms. All use the same fresh evaluation fires.

 The primary comparison, fixed in advance, is the mean of local reward minus simulator credit, $0.0390$ ($95\%$ paired-$t$, seven degrees of freedom: $[0.0212,0.0568]$);  a paired bootstrap over seeds gives $[0.0256,0.0531]$.  The secondary comparison, local reward minus shared reward, has mean $-0.0173$, with paired-$t$ interval $[-0.0247,-0.0098]$ and bootstrap $[-0.0229,-0.0113]$.  Simulator credit beats the local reward, and the local reward beats the shared reward, in every seed pair.

\paragraph[Transfer under earlier rules.]{Transfer under earlier rules.}
 For each feedback signal, eight frozen policies trained in S1 take their most likely action on fire seeds 10000--10029. S2 increases ignition radius from two to four; S3 has three ignitions; S4 reduces aerial water to four and raises no-fly probability to 0.20; S5 has  four hand crews and four aircraft.  The two reference policies are no suppression and a scripted anchor-and-progress heuristic that knows each agent's type and sees the true fire and all positions, using only the main suppression action or none.
\begin{figure}[t]
\centering
\includegraphics[width=\linewidth]{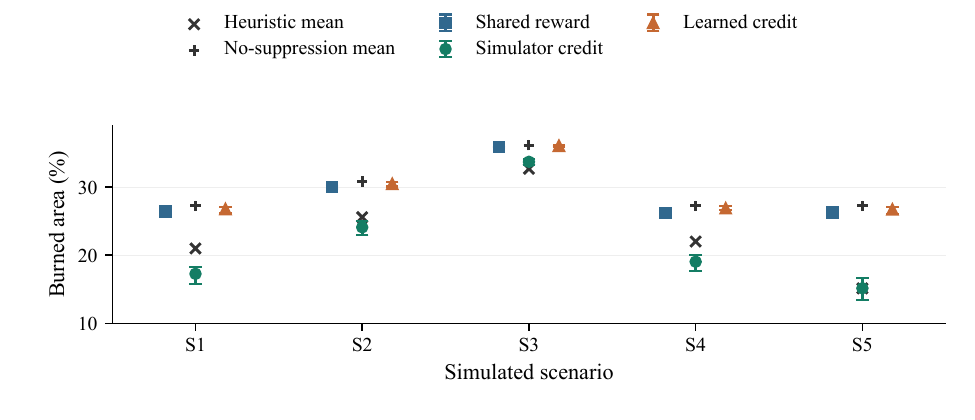}
\caption[Scenario transfer.]{Scenario transfer. Colored symbols show IQMs and  $95\%$ bootstrap intervals over training seeds; black markers show reference means.  All trained policies are the same frozen checkpoints, trained in S1 under earlier rules. S2: larger initial fire; S3: multiple ignitions; S4: tighter aerial constraints;  S5: more aircraft.}
\label{fig:generalization}
\end{figure}
Simulator credit  has the lowest IQM among the trained policies in all five scenarios and beats the heuristic mean in S1, S2 and S4; the heuristic is stronger in S3 and S5. In S1 its IQM is $0.1728$ $[0.1584,0.1825]$, versus shared $0.2643$ $[0.2593,0.2677]$ and learned $0.2680$ $[0.2614,0.2716]$.  Simulator credit wins all eight seed pairs against the shared reward. S2--S5 simulator scores are $0.241$, $0.337$, $0.191$, $0.152$, versus learned scores $0.305$, $0.361$, $0.269$, $0.267$;  in all five scenarios, the paired interval for simulator credit minus learned credit excludes zero.
\section{Critic tuning and actor-logit control}
\label{app:supplement}
\begin{figure}[t]
\centering
\includegraphics[width=\linewidth]{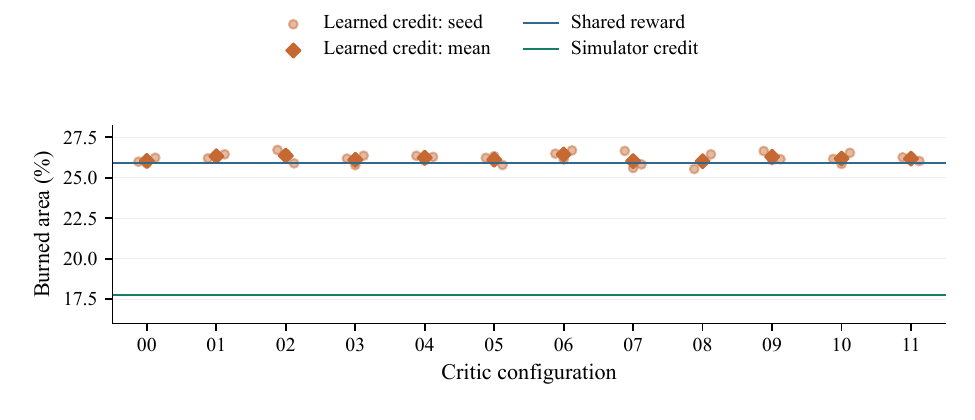}
\caption[Critic sweep under earlier rules.]{Critic sweep for learned credit under earlier rules: twelve configurations, three training seeds each. Each point is a seed's  mean final burn score over its last ten training updates; diamonds average the three seeds. Dashed references use the same metric and seeds 1--3 from the main sweep.  These summarize training; they are neither scores of the tuned policies on new fires nor uncertainty intervals.}
\label{fig:tuning}
\end{figure}

 The critic sweep under earlier rules uses seeds 1--3 and twelve configurations: base; learning rates $10^{-3}$ or $2\times10^{-4}$; width 256; TD($\lambda$) with $\lambda=0.8$ or 1; a target critic; and five combinations.  The target critic copies the critic every 200 updates. All 36 runs complete the full budget.  Means over three seeds and the last ten training updates range from $0.2602$ to $0.2644$, versus shared $0.2594$ and simulator $0.1774$.  The tuned critic (a target critic with TD($\lambda$) and $\lambda=1$; number 08 in Figure~\ref{fig:tuning}) has the lowest mean in the sweep.
 The six runs without the clamp under earlier rules remove only the bound on the actor's logits in sampling, optimization and evaluation; PPO ratio and gradient clipping remain. All finish the full budget. Group means change from $0.2641$ to $0.2683$ for informed critics and $0.2667$ to $0.2683$ for zero-input critics.
\begin{table}[!htbp]
\centering
\small
\setlength{\tabcolsep}{4pt}
\renewcommand{\arraystretch}{1.12}
\begin{tabular}{lrrrr}
\toprule
Critic inputs & Seed & Clamped & No clamp & Difference \\
\midrule
Informed & 1 & $0.259793$ & $0.268792$ & $+0.008999$ \\
Zero-input & 1 & $0.259793$ & $0.270819$ & $+0.011025$ \\
Informed & 2 & $0.269488$ & $0.267045$ & $-0.002443$ \\
Zero-input & 2 & $0.267701$ & $0.267047$ & $-0.000654$ \\
Informed & 3 & $0.263088$ & $0.268929$ & $+0.005841$ \\
Zero-input & 3 & $0.272616$ & $0.266965$ & $-0.005651$ \\
\bottomrule
\end{tabular}
\caption[Paired runs with and without the logit clamp under earlier rules.]{Six paired runs with and without the clamp on the actor's logits, under earlier rules.  Entries are mean burn scores on fires 10000--10029; positive differences favor the clamped actor. PPO ratio clipping and gradient clipping remain enabled. The three seeds per condition are shown separately.}
\label{tab:unclipped-seeds}
\end{table}

\begin{figure}[!htbp]
\centering
\includegraphics[width=\linewidth]{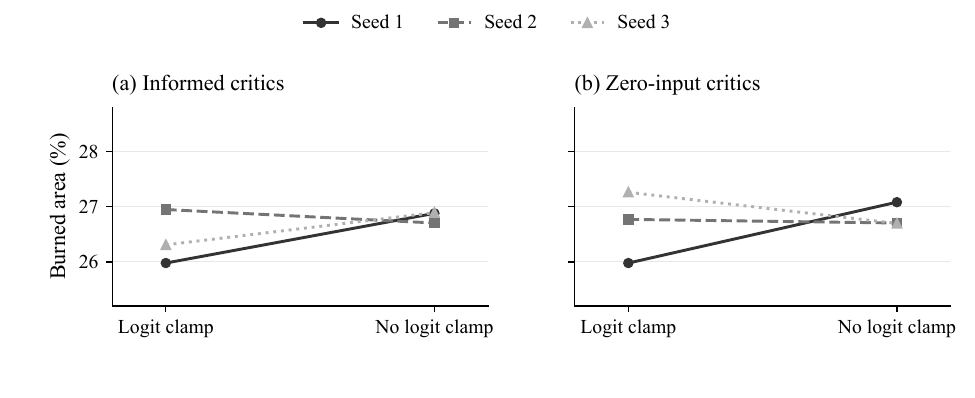}
\caption[Removing the logit clamp under earlier rules.]{Removing the logit clamp under earlier rules. Each line joins  one training seed's mean burn score on thirty fixed fires with and without the clamp. Across these six runs,  the changes for the three informed and three zero-input critics have mixed signs, and the group mean burn score is higher without the clamp in each condition.}
\label{fig:credit-controls}
\end{figure}

\section{Randomness and intervention controls}
\label{app:offline}
 Eight actors trained with simulator credit under earlier rules each play four stochastic episodes in S1 (seeds 40000--40003). Uniform reservoir sampling  selects, in each episode, one context after step 10 with an accepted placement; all 2,880 eligible steps qualify.  The resulting 256 cases, one per agent and context, include 195 with a nonempty placement and 180 whose deletion changes the combined fields. Snapshots fix tactical state, movement, acceptance  and the states of all random number generators.

For raw reward $R_e$, placements $P$,  placements $P_{-i}$ without agent $i$'s, and independent random draws $k,l$ for the fire, compute the credit with common random numbers (draw $k$ in both branches, the coupled case) and with independent draws $k$ and $l$:
\begin{equation}
D^{\rm coupled}_{ei}=R_e(P;k)-R_e(P_{-i};k),\quad
D^{\rm independent}_{ei}=R_e(P;k)-R_e(P_{-i};l).
\end{equation}
Each context uses 256 independent pairs of draws, shared across agents.  Both credits have the same expectation; their variances differ because the two branches covary under shared draws. Dynamics stay float32 and burned fields are summed in float64.  With the random draws used in training, all branches match the serial fire updates exactly, and all credits match those computed during training within $8.95\times10^{-9}$.
\begin{table}[!htbp]
\centering
\small
\begin{tabular}{lrrrr}
\toprule
 Rows included & Count &  Shared draws &  Independent draws & Variance ratio \\
\midrule
All & 256 & $2.08\times10^{-11}$ & $1.94\times10^{-8}$ & $934\;[675,1416]$ \\
Accepted placement & 195 & $2.73\times10^{-11}$ & $1.94\times10^{-8}$ & $709\;[537,1055]$ \\
Deletion changes fields & 180 & $2.96\times10^{-11}$ & $2.00\times10^{-8}$ & $677\;[503,1013]$ \\
\bottomrule
\end{tabular}
\caption[Variance of simulator credit with shared and independent draws.]{Variance of the credit within each context and agent, averaged over cases, in squared raw reward and estimated from 256 pairs of draws.  Ratios divide the averaged variances, because the ratio for a single case can be undefined. Brackets are descriptive 95\% intervals from 2,000 nested bootstrap draws: resample whole actors with their four fixed episode seeds,  then pairs of draws jointly for all agents within each context. Intervals condition on these fire seeds and selected contexts.}
\label{tab:coupling}
\end{table}
 Credit with independent draws is negative in $30{,}749/65{,}536$ samples ($46.92\%$);  credit with shared draws is never negative. Figure~\ref{fig:repaired-learning}c uses the full empirical distribution,  with equal weight on every context, draw and agent.  The variance ratio averages the variances within each context and agent, including 196 cases whose sampled variance with shared draws is zero.

\paragraph{Choice of removal.}
 At the realized draw of each case, replay from the full state replaces agent $i$'s action by $(m_i,\mathrm{none})$, the same move without suppression, or by $(\mathrm{stay},\mathrm{none})$, staying in place without suppression; teammates' actions are kept and the rules are applied again.  Compared with deleting the placement, both give exactly the same immediate credit in all 256 cases (tolerance fixed in advance: $10^{-7}$)  and change no teammate's placement.  Each changes the water in 15 cases and the refill timer in three; staying in place also changes positions in 243. These measurements concern the immediate reward, with tactical differences recorded separately.
\section{Audit protocol and current-rule results}
\label{app:audit}\label{app:audit-protocol}\label{app:completed-controls}
 We audit four informed critics from the main comparison (seeds 1, 4, 7, 8), selected because their final actors are not saturated (Section~\ref{sec:method}).  Ten critics trained under earlier rules test whether the results are consistent across training runs and how they depend on the critic's inputs.  Each critic is audited with its own actor, inputs and normalizers after the full training budget. Informed critics receive 109 bounded features including positions, types, resources, timers, previous movement, activity, time and dynamics indicators;  zero-input critics keep this architecture but receive zeros in place of these tactical features.  Each audit uses the rules under which its critic was trained.

\paragraph{Sampling and uncertainty.}
 For each checkpoint, uniform reservoir sampling selects one context at or after step 10 in each of eight stochastic episodes (seeds 40000--40007),  and each action is replayed to the end of the episode under sixteen draws shared by all actions, with discount $\gamma=.99$.  Under the current rules, the selected steps are 96, 63, 89, 55, 82, 92, 65, 18. Contexts are independent within a checkpoint; checkpoints are analyzed separately.  Metrics weight agents and contexts equally and weight actions by the saved policy's probabilities.  Intervals resample whole episodes 10,000 times, conditional on the stored replay means. Stored PopArt parameters and  the final running standard deviation of the reward convert predictions to raw units.

With centered draws $Z_{ei,m}(b)$ from Equation~\eqref{eq:audit-replay-reference},  the Monte Carlo standard error of each replay credit is
\begin{equation}
\widehat{\mathrm{SE}}_{ei}(b)=\sqrt{\frac{\sum_m[Z_{ei,m}(b)-\widehat A_{ei}(b)]^2}{M(M-1)}}.
\end{equation}
 For the critic's error minus the error of zero credit, define $D_{e,m}=N^{-1}\sum_{i,b}p_{ei}(b)[(A^c_{ei}(b))^2-2A^c_{ei}(b)Z_{ei,m}(b)]$.  Its Monte Carlo standard error (MCSE) is $E^{-1}\sqrt{\sum_e\widehat{\mathrm{Var}}_m(D_{e,m})/M}$.  This keeps the covariance across actions and agents and cancels the squared reference term. RMSE ratios below are point estimates; inferential intervals use paired error differences.
\begin{table}[!htbp]
\centering
\small
\setlength{\tabcolsep}{4pt}
\begin{tabular}{rrrrrr}
\toprule
Seed & Factual $R^2$ & Factual RMSE & Credit RMSE & Zero RMSE & Replay MCSE \\
\midrule
1 & 0.954 & 11.716 & 1.264 & 0.143 & 0.149 \\
4 & 0.911 & 15.910 & 1.789 & 0.120 & 0.102 \\
7 & 0.950 & 11.617 & 1.120 & 0.072 & 0.115 \\
8 & 0.935 & 13.395 & 2.087 & 0.171 & 0.159 \\
\bottomrule
\end{tabular}
\caption[Return fit and credit error under the current rules.]{Return fit and credit error of the four critics audited under the current rules. All RMSE and replay MCSE columns are multiplied by $10^3$.  Credit and zero errors weight contexts and agents equally and weight actions by the saved policy. Replay MCSE is  the square root of the variance of the sixteen-draw mean credit, with the same weights.}
\label{tab:current-audit-calibration}
\end{table}
\begin{table}[!htbp]
\centering
\small
\setlength{\tabcolsep}{4pt}
\begin{tabular}{rlrr}
\toprule
Seed & $10^6\Delta$ [$95\%$ interval] & Separate MCSE & RMSE / zero \\
\midrule
1 & 1.577 [0.629, 2.843] & 0.027 & 8.81 \\
4 & 3.185 [1.308, 5.720] & 0.045 & 14.85 \\
7 & 1.248 [0.234, 2.552] & 0.043 & 15.59 \\
8 & 4.324 [1.381, 7.858] & 0.071 & 12.20 \\
\bottomrule
\end{tabular}
\caption[Critic error against zero credit under the current rules.]{Raw credit error of the critic minus the error of zero credit, and the ratio of their RMSEs, under the current rules.  The MCSE column is also multiplied by $10^6$.  Intervals reuse the 10,000 bootstrap samples of complete episodes fixed in advance, conditional on each checkpoint and its stored replay means.  RMSE ratios are point estimates against the sixteen-draw replay reference, without correction for its noise.}
\label{tab:current-audit-difference}
\end{table}
\begin{table}[!htbp]
\centering
\small
\setlength{\tabcolsep}{4pt}
\begin{tabular}{rrrrrr}
\toprule
Seed & Unique pairs & Coverage (\%) & Contexts & Reversals (\%) & Critic agreement (\%) \\
\midrule
1 & 823 & 6.77 & 6/8 & 40.29 & 51.60 \\
4 & 732 & 6.02 & 7/8 & 24.41 & 51.36 \\
7 & 313 & 2.57 & 3/8 & 60.52 & 56.55 \\
8 & 545 & 4.48 & 5/8 & 24.49 & 60.90 \\
\bottomrule
\end{tabular}
\caption[Screening of action pairs under the current rules.]{Screening of action pairs with independent halves of the draws, under the current rules. Each checkpoint has 12,160  possible unique pairs of actions over contexts and agents.  One half of eight draws selects pairs and the other evaluates them, then the halves swap roles. Coverage counts each pair once;  reversals and critic agreement average first within each episode, as fixed in advance, across both folds.}
\label{tab:current-audit-direction}
\label{tab:current-audit-pairs}
\end{table}
\begin{table}[!htbp]
\centering
\small
\setlength{\tabcolsep}{4pt}
\begin{tabular}{rrlrr}
\toprule
Seed & Spearman & $95\%$ episode interval & Defined contexts & Defined agent cases \\
\midrule
1 & +0.001 & [-0.049, 0.050] & 6/8 & 33/64 \\
4 & -0.036 & [-0.138, 0.062] & 7/8 & 45/64 \\
7 & +0.067 & withheld & 3/8 & 17/64 \\
8 & -0.025 & [-0.090, 0.038] & 6/8 & 35/64 \\
\bottomrule
\end{tabular}
\caption[Rank agreement within agents under the current rules.]{Spearman correlation between the critic's and the replay's rankings of each agent's actions under the current rules, averaged first within each episode and then over defined episodes.  Flat replay cases have no defined rank correlation and remain in the reported denominators. The interval is withheld when fewer than five episodes define the metric.}
\label{tab:current-audit-ranks}
\end{table}

\paragraph{Ranking and selectivity.}
For each action pair, set $D_m=G_m(a)-G_m(b)$.  One half of the draws (eight draws) selects pairs with $|\bar D|>t_{0.95,7}s_D/\sqrt8+10^{-14}$  and whose means over its two sets of four draws have the same nonzero sign; the other half checks the direction, and then the halves swap roles. Held-out ties use the same $10^{-14}$ tolerance, and a tied critic gets half credit. Coverage counts unique pairs; direction agreement excludes held-out ties and gives tied critic predictions half credit.  Summaries over folds and agents average within episodes, then across the episodes represented; reversals retain all selected pairs.  A flat case, in which all twenty actions have the same replay credit, has no defined Spearman correlation; intervals require at least five contexts in which it is defined.  The numbers of flat cases (all twenty returns equal within each draw) are $31/64,19/64,47/64,29/64$ for  seeds 1, 4, 7, 8 under the current rules.  The median Spearman correlations between agents' rankings, over each critic's 28 pairs of agents in each of eight contexts, are $0.986,0.988,0.985,0.980$.  This check uses the critics alone; the precision of the replay reference for the direction of credit is reported next to it.
\label{app:agent-patterns}
\paragraph[Precision of the replay reference.]{Precision of the replay reference.}
 For seeds 1, 4, 7 and 8 under the current rules, the policy-weighted cosines between the replay credits from two independent halves of eight draws are $0.002$, $0.147$, $-0.449$ and $0.104$.  Cosines between the frozen actor-score gradients from critic credit and from replay credit range from $-0.109$ to $+0.662$,  and all Monte Carlo intervals for their inner products span zero (Table~\ref{tab:current-gradients}). Action-pair screening is reported in Table~\ref{tab:current-audit-pairs}.  These checks accompany the calibration results; they do not resolve the order of actions.

\section[Earlier audits and actor saturation]{Earlier audits and actor saturation}
\label{app:earlier-controls}
 The ten critics trained under earlier rules are three base critics, the tuned critic, three informed critics and three capacity-matched zero-input critics.  All ten have a larger error than zero credit under both policy and uniform weights.  Base critics 1 and 3 have actors that are not saturated on the checked trajectories, and their raw RMSE is $9.09$ and $7.01$ times  that of zero credit.  The actors of base critic 2, the tuned critic and all three informed critics are fully saturated on their checked trajectories.  Pooled over the three earlier informed critics, the rank correlation between agents is $0.997$.
\begin{table}[!htbp]
\centering\small\setlength{\tabcolsep}{3pt}
\begin{tabular}{lrrrlr}
\toprule
Critic & $R^2$ & Credit & Zero & $10^6\Delta$ [$95\%$] & MCSE \\
\midrule
Base seed 1 & $0.938$ & $1.758$ & $0.193$ & $3.0545$ $[1.7178,4.4575]$ & $0.0561$ \\
Base seed 2 & $0.808$ & $0.516$ & $0.203$ & $0.2252$ $[0.1416,0.3106]$ & $0.0094$ \\
Base seed 3 & $0.890$ & $1.340$ & $0.191$ & $1.7604$ $[0.5492,3.5325]$ & $0.0666$ \\
 Tuned seed 1 & $0.902$ & $0.880$ & $0.110$ & $0.7631$ $[0.4884,1.1470]$ & $0.0110$ \\
Informed seed 1 & $0.957$ & $0.354$ & $0.178$ & $0.0938$ $[0.0689,0.1148]$ & $0.0080$ \\
Informed seed 2 & $0.970$ & $0.374$ & $0.088$ & $0.1326$ $[0.1164,0.1487]$ & $0.0063$ \\
Informed seed 3 & $0.962$ & $0.447$ & $0.237$ & $0.1442$ $[0.0769,0.2240]$ & $0.0116$ \\
Zero-input seed 1 & $0.879$ & $0.554$ & $0.189$ & $0.2709$ $[0.1370,0.4254]$ & $0.0040$ \\
Zero-input seed 2 & $0.907$ & $1.189$ & $0.410$ & $1.2457$ $[0.3814,2.4101]$ & $0.0691$ \\
Zero-input seed 3 & $0.755$ & $0.726$ & $0.177$ & $0.4955$ $[0.1159,0.9879]$ & $0.0029$ \\
\bottomrule
\end{tabular}
\caption[Audits of the ten critics trained under earlier rules.]{Audits of the ten critics trained under earlier rules, with eight contexts and sixteen draws each.  The credit and zero columns are RMSE in $10^{-3}$ return;  the paired differences and their MCSE are in $10^{-6}$ return squared. Intervals resample episodes conditional on stored means. All ten checkpoints remain represented.}
\label{tab:earlier-audit-summary}
\end{table}

\paragraph{Final-actor diagnostics.}
\label{app:v2-saturation}
Four trajectories per actor  (two with sampled actions, two with the most likely action) and  one gradient context after step 10 per trajectory test whether the final policy can still change.  For learned credit under the current rules, the checks use stochastic seeds 40000--40001 and evaluation seeds 2026092400--2026092401; all deterministic outcomes reproduce saved evaluations. Table~\ref{tab:v2-saturation} identifies the non-saturated subset,  whose mean burn score is $0.2693$, versus  $0.2670$ for the paired seeds trained with the shared reward. Earlier MAPPO checks are in Table~\ref{tab:headline-saturation}; all 16 actors have  nonzero gradients of the sampled negative log likelihood (NLL) and of the entropy.
\begin{table}[!htbp]
\centering\small\setlength{\tabcolsep}{3pt}
\begin{tabular}{rrrrrrrr}
\toprule
& \multicolumn{3}{c}{Learned credit (current rules)} & \multicolumn{4}{c}{Earlier MAPPO} \\
Seed & Saturated & NLL & Entropy & Shared sat.\% & Sim. sat.\% & Shared NLL & Sim. NLL \\
\midrule
1 & $0/3200$ & $1.107$ & $2.762$ & $0.00$ & $0.28$ & $2.89$ & $2.04$ \\
2 & $3200/3200$ & $0.000$ & $0.000$ & $0.00$ & $0.46$ & $3.31$ & $3.05$ \\
3 & $3200/3200$ & $0.000$ & $0.000$ & $0.00$ & $0.44$ & $4.63$ & $2.07$ \\
4 & $0/3200$ & $5.945$ & $3.697$ & $0.00$ & $0.00$ & $3.66$ & $2.18$ \\
5 & $3200/3200$ & $0.000$ & $0.000$ & $0.00$ & $1.71$ & $4.29$ & $3.98$ \\
6 & $3200/3200$ & $0.000$ & $0.000$ & $0.00$ & $3.66$ & $3.74$ & $2.49$ \\
7 & $0/3200$ & $0.934$ & $3.168$ & $1.03$ & $0.09$ & $3.11$ & $2.76$ \\
8 & $0/3200$ & $3.775$ & $4.177$ & $0.00$ & $0.81$ & $3.66$ & $4.28$ \\
\bottomrule
\end{tabular}
\caption[Checks of the final actors.]{Final-actor checks: four trajectories and four gradient contexts per actor.  Under the current rules, saturation counts are out of 3,200 decisions; under earlier rules, percentages cover 2,568--3,200 decisions.  The NLL and entropy columns give the smallest norm, over the checked contexts, of the gradient with respect to the actor's parameters. Decisions within trajectories are dependent.}
\label{tab:v2-saturation}\label{tab:headline-saturation}
\end{table}
 All eight actors trained with the local reward have nonzero gradients on at least one checked context. Fully saturated decisions out of 3,200 are $250,122,112,186,2837,2053,106,330$ for seeds 1--8.  On the six seeds of the local reward with fewer than $10.4\%$ saturated decisions (1--4, 7--8),  the mean burn score is $0.2475$ versus $0.2034$ for  the paired policies trained with simulator credit, which win all six pairs (Table~\ref{tab:repaired-seeds}). This post hoc subset retains the performance gap after excluding seeds 5 and 6. Every training seed remains in the primary comparison.

\section[Counterfactual targets with common random numbers and finite-data coverage]{Counterfactual targets with common random numbers and finite-data coverage}
\label{sec:matched-credit-theory}
These standard identities specify the measurement framework. Use the complete context, frozen policy and remaining horizon from Equations~\eqref{eq:main-target}--\eqref{eq:audit-replay-reference}; teammates' actions are fixed at the intervention,  and afterwards all actors act on the observations of their own branch.

\paragraph[Replay estimator with common random numbers.]{Replay estimator with common random numbers.}
For fixed $z_t,\mathbf a,\pi$, let $Z_{i,m}$ be the centered factual-action return under draw $m$ and $\widehat A=M^{-1}\sum_mZ_{i,m}$.  If every branch has the correct marginal distribution and finite variance, and the $M$ draws are independent, linearity and independence give
\begin{equation}
\mathbb E[\widehat A]=A_i^\pi(a_i),\qquad
\operatorname{Var}(\widehat A)=\operatorname{Var}(Z_{i,1})/M.
\end{equation}
 Within one draw, all branches share the same random numbers. For two returns with fixed marginals,
$\operatorname{Var}(G-G')=\operatorname{Var}(G)+\operatorname{Var}(G')-2\operatorname{Cov}(G,G')$; positive covariance reduces difference variance relative to independent draws.

\paragraph{Finite-data counterfactual non-identification.}
Let $L(q)=\sum_{k=1}^n(q(z_k,\mathbf a_k)-y_k)^2$ have fixed regression labels and evaluated inputs $\mathcal D=\{(z_k,\mathbf a_k)\}_{k=1}^n$. For factual $(z,\mathbf a)\in\mathcal D$, consider an unobserved alternative $u=(z,(\mathbf a_{-i},b))\notin\mathcal D$, where $b\ne a_i$ and $p_b=\pi_i(b\mid h_i)>0$. If the function class permits $q_\lambda=q+\lambda\mathbf1_{\{u\}}$, then
\begin{equation}
L(q_\lambda)=L(q),\qquad
A_i[q_\lambda](z,\mathbf a)=A_i[q](z,\mathbf a)-p_b\lambda.
\end{equation}
Here $A_i[q]$ is  the contrast of Equation~\eqref{eq:main-target} for the action taken, with action values replaced by $q$. The perturbation vanishes on observed inputs and changes only the baseline's term for $b$. This holds for an unrestricted tabular class; regularization and structure can constrain unseen values. Full action support and zero population error against the correct conditional-mean target identify the contrasts.

\paragraph{Tabular action coverage.}
Fix a context $z$ and anchor joint action $\mathbf a$ independently of $n$ i.i.d. draws from a stationary behavior distribution $\mu(\cdot\mid z)$. Let $U_i$ count unvisited focal alternatives with positive policy probability. An alternative of probability $p$ is absent with probability $(1-p)^n$, giving
\begin{equation}
\mathbb E[U_i]=\sum_{\substack{b\ne a_i\\\pi_i(b\mid h_i)>0}}
\left(1-\mu((\mathbf a_{-i},b)\mid z)\right)^n.
\label{eq:flare-general-occupancy}
\end{equation}
Under independent uniform $K$-action sampling by all $N$ agents in both $\mu$ and $\pi$, this is $(K-1)(1-K^{-N})^n$ per agent and $N(K-1)(1-K^{-N})^n$ across the distinct single-agent alternatives. Here $n$ counts visits to the same complete context. The unseen fraction grows with $N$ at fixed $n,K$; $20^8=25.6\times10^9$ counts joint actions. Neural generalization additionally depends on structure and factorization; this occupancy identity motivates a coverage intervention.

\section{Scale sensitivity and frozen actor-score gradients}
\label{app:scale-sensitivity}
\paragraph{RMS matching and held-out fitting.}
Equation~\eqref{eq:rms-credit-error}  removes an overall inflation of the critic's credit by rescaling it to the root mean square (RMS) of the replay reference. It differs from PPO's centering and standardization of sampled batch advantages. Separately, define $W_e[f]=N^{-1}\sum_{i,b}p_{ei}(b)f_{ei}(b)$ and fit one nonnegative scalar on seven contexts, testing on the eighth:
\begin{align}
 \alpha_{-e}&=\max\!\left\{0,\frac{\sum_{d\ne e}W_d[A^c\widehat A]}{\sum_{d\ne e}W_d[(A^c)^2]}\right\},\nonumber\\
 \Delta^{\mathrm{scale}}&=\frac18\sum_{e=1}^{8}W_e\!\left[\alpha_{-e}^{2}(A^c)^2-2\alpha_{-e}A^c\widehat A\right].
 \label{eq:scale-heldout}
\end{align}
 Each bootstrap resample of whole episodes refits every fold, excluding all copies of the episode held out in that fold. Intervals condition on saved replay draws. Table~\ref{tab:current-scale}  reports the raw, RMS-matched and fitted errors under the current rules separately.  Under earlier rules, the RMS-matched error is $1.797$--$2.235$ times  that of zero credit; held-out fitting gives $1.0000$--$1.0073$,  and every interval for the fitted error minus the error of zero credit includes or touches zero.  For the earlier critics, cosines between critic and replay credit range from $-0.118$ to $+0.101$,  cosines between the two halves of the draws from $-0.314$ to $+0.573$, and estimated noise ratios from $0.535$ to $1.505$.  The noise ratio is the Monte Carlo variance divided by the squared size of the replay reference. Once the scale is removed, these errors depend on the order of the actions' credit, which a reference built from sixteen draws is too noisy to grade (Section~\ref{sec:audit-results}), so we report them as unresolved.
\begin{table}[!htbp]
\centering
\small
\setlength{\tabcolsep}{3pt}
\begin{tabular}{rrrrlr}
\toprule
Seed & Raw MSE / zero & Rescaled MSE / zero & Fitted MSE / zero & $10^{10}\Delta^{\mathrm{scale}}$ [$95\%$] & Cosine \\
\midrule
1 & 77.64 & 1.818 & 1.005 & $1.1\;[-0.00338, 3.12]$ & 0.091 \\
4 & 220.48 & 1.619 & 0.994 & $-0.834\;[-13.2, 6.13]$ & 0.190 \\
7 & 243.19 & 1.440 & 1.019 & $0.973\;[-0.654, 3.99]$ & 0.280 \\
8 & 148.79 & 1.803 & 1.044 & $12.8\;[0, 52.4]$ & 0.098 \\
\bottomrule
\end{tabular}
\par\medskip
\begin{tabular}{rrrr}
\toprule
Seed & RMS multiplier & Minimum $\alpha_{-e}$ & Maximum $\alpha_{-e}$ \\
\midrule
1 & 0.1130 & 0.0004 & 0.0129 \\
4 & 0.0666 & 0.0042 & 0.0151 \\
7 & 0.0631 & 0.0029 & 0.0238 \\
8 & 0.0816 & 0 & 0.0187 \\
\bottomrule
\end{tabular}
\caption[Raw and rescaled credit errors under the current rules.]{Raw and rescaled credit errors under the current rules; the rescaled checks are unresolved (Section~\ref{sec:audit-results}). ``Raw MSE / zero'', ``Rescaled MSE / zero'' and ``Fitted MSE / zero''  divide the mean squared error (MSE) of the raw credit, of the credit rescaled to the reference's RMS and of the held-out fit by the MSE of zero credit.  Cosine compares critic and replay credit.  Agreement between independent halves is reported in Appendix~\ref{app:audit}. Episode intervals refit the common scale  in held-out folds that exclude every copy of the held-out episode. The lower panel gives  the RMS multiplier and the smallest and largest fitted multipliers across all eight held-out folds.}
\label{tab:current-scale}
\end{table}
\paragraph{Frozen actor-score gradients.}
 Using the saved actors trained under the current rules and the exact observations from replay, define
\begin{equation}
 g(A)=\frac{1}{8N}\sum_{e=1}^{8}\sum_{i=1}^{N}\sum_{b=1}^{20}
 p_{ei}(b)A_{ei}(b)\nabla_\theta\log\pi_\theta(b\mid o_{ei}).
 \label{eq:frozen-score-gradient}
\end{equation}
Credits and probabilities are held fixed; derivatives include all actor parameters. This is a frozen score gradient,  distinct from a full update with PPO, GAE, the entropy term and Adam.  For the replay gradient $g^r_{e,m}$ of context $e$ and draw $m$ and the critic gradient $g^c=g(A^c)$, the MCSE of their inner product is
\begin{equation}
 \widehat{\mathrm{SE}}_{\rm MC}
 =\sqrt{\frac{1}{8^2 M}\sum_{e=1}^{8}
 \widehat{\operatorname{Var}}_m\!\left[g^c\cdot g^r_{e,m}\right]}.
\end{equation}
Table~\ref{tab:current-gradients}  uses Monte Carlo intervals from a normal approximation; bootstrap uncertainty over whole contexts is separate. Zero norms give undefined cosines.  In all ten contexts with a zero replay gradient, all actions have identical returns within every sampled draw.  The agreement between halves and the Monte Carlo intervals measure how precisely the reference can show the direction of credit.
\begin{table}[!htbp]
\centering
\small
\setlength{\tabcolsep}{4pt}
\begin{tabular}{rrrrl}
\toprule
Seed & Gradient cosine & Half cosine & Zero contexts & $10^7$ inner product [$95\%$ MC] \\
\midrule
1 & -0.012 & -0.151 & 2/8 & $-0.033\;[-3.317, 3.251]$ \\
4 & 0.662 & 0.410 & 1/8 & $9.135\;[-8.197, 26.466]$ \\
7 & 0.652 & -0.895 & 5/8 & $1.369\;[-0.874, 3.611]$ \\
8 & -0.109 & 0.287 & 2/8 & $-1.463\;[-5.996, 3.070]$ \\
\bottomrule
\end{tabular}
\caption[Frozen actor-score gradients under the current rules.]{Frozen actor-score gradients for the four critics audited under the current rules.  Cosines use the gradients averaged over contexts.  All four Monte Carlo intervals for the inner product of critic and replay gradients span zero. ``Zero contexts''  counts contexts whose sampled mean replay gradient is exactly zero, so their cosines are undefined.}
\label{tab:current-gradients}
\end{table}
\section{Supplementary visual diagnostics}
\label{app:visuals}

\begin{figure}[!htbp]
\centering
\includegraphics[width=\linewidth]{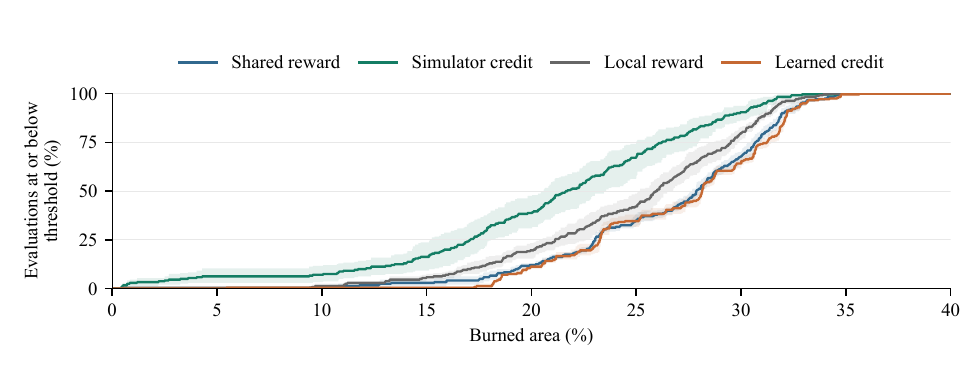}
\caption[Performance profiles.]{Empirical performance profiles for  the four feedback signals.  At each burn score threshold, the curve gives the fraction of the $8\times30$ recorded evaluations at or below that threshold. Shading gives pointwise descriptive $95\%$ intervals from  10,000 resamples of whole training seeds, retaining all thirty fixed fires within each seed.  The curve describes all 240 outcomes of each signal, and the resampling keeps them grouped by training seed.}
\label{fig:performance-profile}
\end{figure}
\begin{figure}[!htbp]
\centering
\includegraphics[width=\linewidth]{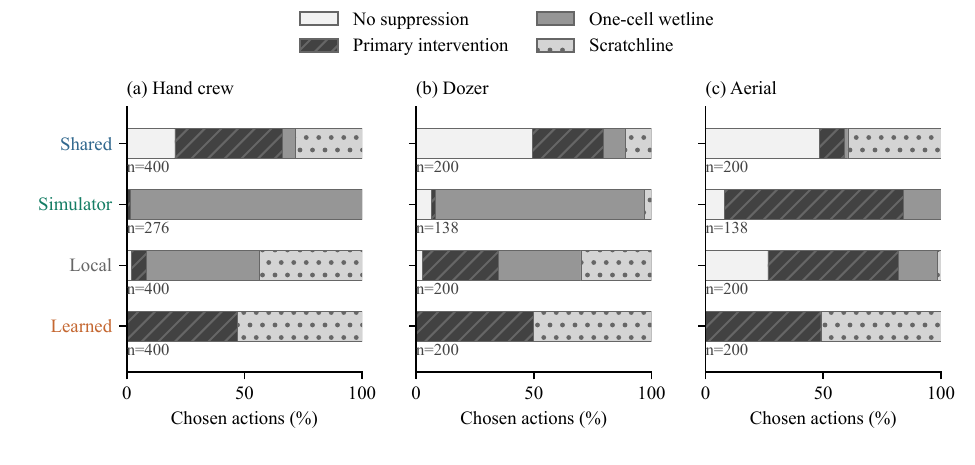}
\caption[Suppression choices by agent type.]{Chosen suppression primitives by agent type  for the four feedback signals, using training seed 1 and fire 2026092400.  The episodes for the shared reward, simulator credit and learned credit are those of Figure~\ref{fig:fire-paths};  the saved episode with the local reward reproduces its recorded evaluation. Each row sums to $100\%$ of $n$ decisions; movement is omitted.  The main suppression action is a fireline for ground crews and a patch drop for aircraft.  With simulator credit the episode ends after 69 steps, and with the other signals after 100. Ground wetline is unanchored and has no modeled water cost.}
\label{fig:action-mix}
\end{figure}
\begin{figure}[!htbp]
\centering
\includegraphics[width=\linewidth]{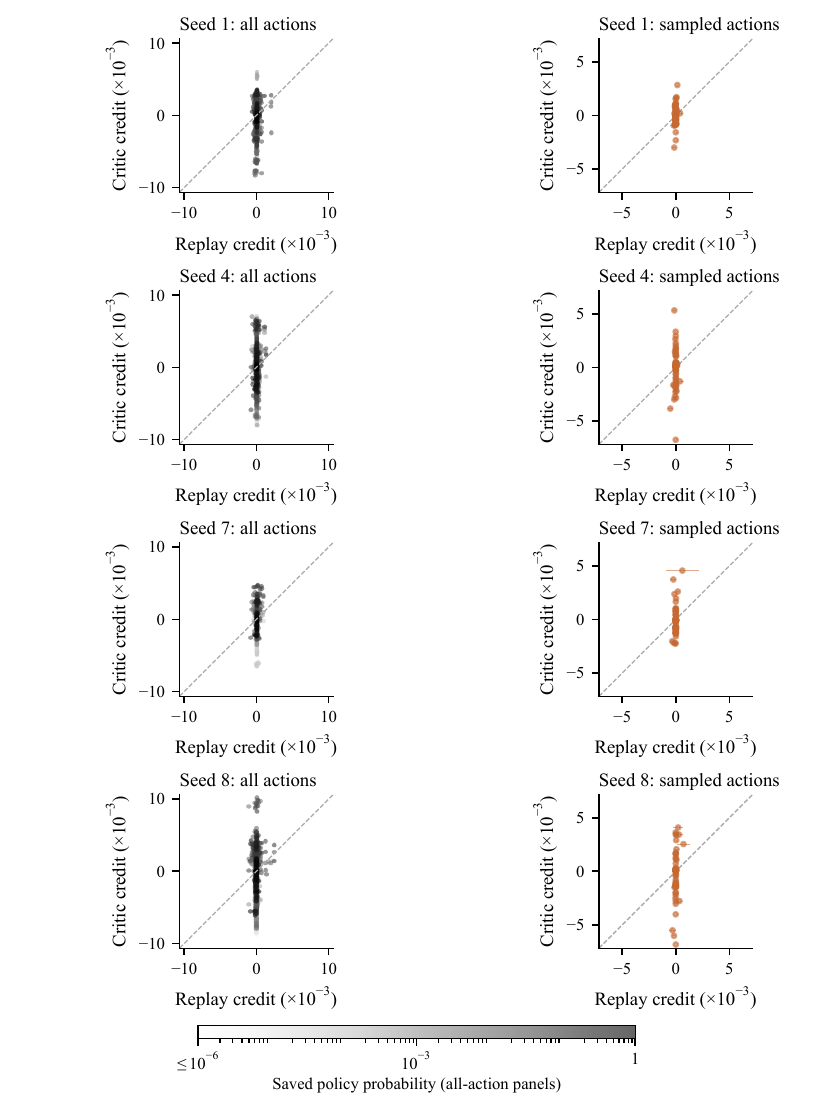}
\caption[Detailed audits under the current rules.]{Detailed audits under the current rules, one row per non-saturated actor and its informed critic (seeds 1, 4, 7 and 8). Left: all twenty actions, $1{,}280$ dependent points per checkpoint,  shaded by the saved policy's probability. Right: the 64 sampled actions, with horizontal $\pm1$ Monte Carlo standard errors.  Each column uses the same axes; credits are centered by the policy, and the main error metric weights actions by the saved policy.  Replay estimates use sixteen continuations to the end of the episode, with draws shared by all actions.}
\label{fig:audit-all-actions}
\end{figure}
\paragraph{Plotting protocol.}
Figure~\ref{fig:repaired-learning}  uses trailing means over 50 updates within each seed and bootstrap bands over whole seeds; the gray band of $\pm1$ point centers on the arithmetic mean of the shared reward. Figure~\ref{fig:informed-audit}  takes its example from seed 1's first stored context (episode 40000, step 96),  uses stable centering for flat cases, and keeps separate errors for the fourteen checkpoints. Illustration captions state the fixed fire, policy seed and episode lengths.

\clearpage
\section[Experiments added in the revision]{Experiments added in the revision}
\label{app:added-experiments}
This appendix reports the experiments added in the revision: retraining under two rule changes (Table~\ref{tab:added-rules}), repairs to learned credit (Table~\ref{tab:added-repairs}) and larger replay budgets for grading critics (Table~\ref{tab:added-replays}). Training and evaluation follow Section~\ref{sec:protocol}: eight seeds unless stated, the same thirty new fires, interquartile means, and $95\%$ intervals from 10,000 paired bootstrap resamples of training seeds. All training rows in Tables~\ref{tab:added-rules} and~\ref{tab:added-repairs} were run in the same round, so they compare directly. For replay labels, each update replays 4 of the 19 untaken actions at each of 128 random agent steps for one step, under the same random draw and with the teammates' actions fixed, and also trains the critic toward $y_b = G_t + (r_b - r_a) + \gamma (v_b - v_a)$, where $G_t$ is its usual target for the action $a$ actually taken, $r_b - r_a$ is the difference in that step's reward, and $v_b - v_a$ is the difference in its values at the two next states.

\begin{table}[!htbp]
\centering\small
\setlength{\tabcolsep}{4pt}
\begin{tabular}{lrrrll}
\toprule
Rules & Shared & Local & Simulator & Lead [$95\%$] & Pairs won \\
\midrule
Current rules & $26.6$ & $25.0$ & $19.3$ & $7.3$ $[5.9,8.5]$ & 8/8, both \\
Limited water & $26.8$ & $25.1$ & $21.7$ & $5.2$ $[4.3,5.6]$ & 8/8, both \\
Firelines from burned ground & $23.1$ & -- & $14.9$ & $8.2$ $[4.9,12.2]$ & 8/8, shared \\
Both changes & $21.2$ & $17.5$ & $11.3$ & $9.9$ $[7.2,12.3]$ & 8/8, both \\
\bottomrule
\end{tabular}
\caption[Retraining under two rule changes.]{Burn score (\%, lower is better) under the current rules and under the two rule changes of Section~\ref{sec:learning-results}. With limited water, each ground crew carries eight wetline uses and cannot place wetline until a refill of five steps ends; the second change lets a fireline also start next to burned ground. Shared, local and simulator are the shared reward, the local reward and simulator credit; the local reward was not trained with the second change alone. The lead is the shared reward's burn score minus that of simulator credit, and the last column gives the seed pairs that simulator credit wins against both other signals or against the shared reward.}
\label{tab:added-rules}
\end{table}

\begin{table}[!htbp]
\centering\small
\setlength{\tabcolsep}{4pt}
\begin{tabular}{lrrrr}
\toprule
Training signal & Seeds & Fail check & Burn score (\%) & Worse than shared \\
\midrule
Shared reward & 8 & 0 & $26.6$ & -- \\
Simulator credit & 8 & 0 & $19.3$ & -- \\
Learned credit, no clamp & 8 & 1 & $27.3$ & 8/8 \\
\quad + replay labels & 8 & 3 & $26.9$ & 8/8 \\
\midrule
\multicolumn{5}{@{}l}{\emph{Three-seed trial (seeds 2, 3 and 7)}} \\
Learned credit, no clamp & 3 & -- & $27.2$ & 3/3 \\
\quad + simple regularizer, no replay & 3 & -- & $27.1$ & 3/3 \\
\quad + one output head per agent type & 3 & -- & $27.2$ & 3/3 \\
\quad + replay labels and output heads & 3 & -- & $26.9$ & 3/3 \\
\bottomrule
\end{tabular}
\caption[Repairs to learned credit.]{Repairs to learned credit under the current rules, all without the clamp on the policy's logits. Burn score is the interquartile mean for eight seeds and the mean for three seeds. Fail check counts actors flagged by the training check, fixed before training, which flags an actor whose actions barely depend on what it observes or that still acts almost at random; the last column counts seed pairs in which learned credit burns more than the shared reward. Replay labels lower the burn score by $0.4$ points ($95\%$ interval $[0.0,0.7]$). The simple regularizer uses $G_t$ as the label for the same untaken actions, with no replay. In the three-seed trial, the policy froze in 1, 2 and 2 of the three runs with the regularizer, the output heads and both repairs, and in none without them. Output heads alone left agents' rankings nearly identical in two of three runs; with replay labels added, the median Spearman correlation between agents' rankings fell from about $0.98$ to between $0.14$ and $0.30$.}
\label{tab:added-repairs}
\end{table}

\begin{table}[!htbp]
\centering\small
\setlength{\tabcolsep}{4pt}
\begin{tabular}{llccc}
\toprule
Replays per action & Moments & Size ratio & Signal-to-noise ratio & Cases with no change \\
\midrule
16 & Original & 9--16 & 0.6--1.2 & 19--47 \\
16 & Earlier & 11--23 & 0.9--1.0 & 3--8 \\
64 & Original & 20--31 & 0.9--1.2 & 5--44 \\
64 & Earlier & 20--42 & 1.1--1.4 & 0--6 \\
256 (seed 8) & Earlier & 34 & 2.0 $[1.1,3.0]$ & 4 \\
\bottomrule
\end{tabular}
\caption[Larger replay budgets for grading critics.]{Grading the four critics of Section~\ref{sec:audit-results} at the original eight moments, seven of them late in the episode, and at eight earlier moments (steps 10 to 48). Ranges cover the four critics. The size ratio is the RMSE of the critic's credit divided by that of giving every action zero credit; replay noise inflates the divisor, so the ratio rises as replays are added. The signal-to-noise ratio is the root mean square of the replay credit divided by its Monte Carlo error, with signal equal to noise at $\sqrt{2}$. The last column counts, among each critic's 64 cases (eight agents at eight moments), those in which none of the 20 actions changes the outcome in any replay. Seed 8, the critic closest to the precision threshold at 64 replays, was replayed 256 times at its earlier moments, where its size ratio is 20 with 16 replays.}
\label{tab:added-replays}
\end{table}

\end{document}